\documentclass[11pt]{article}
\usepackage{graphicx}
\usepackage{amssymb,amsmath,amsfonts}

\def\Re{{\cal R \mskip-4mu \lower.1ex \hbox{\it e}\,}}
\def\Im{{\cal I \mskip-5mu \lower.1ex \hbox{\it m}\,}}
\def\ie{{\it i.e.}}
\def\eg{{\it e.g.}}

\def\etal{{\it et al.}}

\def\sub#1{_{\lower.25ex\hbox{$\scriptstyle#1$}}}
\def\tev{\,{\ifmmode\mathrm {TeV}\else TeV\fi}}
\def\gev{\,{\ifmmode\mathrm {GeV}\else GeV\fi}}
\def\mev{\,{\ifmmode\mathrm {MeV}\else MeV\fi}}
\def\mpl{\ifmmode M_{pl}\else $M_{pl}$\fi}
\def\mpl{\ifmmode \overline M_{Pl}\else $\bar M_{Pl}$\fi}
\def\to{\rightarrow}

\def\subw{_{\rm w}}
\def\mh{\ifmmode m\sbl H \else $m\sbl H$\fi}
\def\mch{\ifmmode m_{H^\pm} \else $m_{H^\pm}$\fi}
\def\mt{\ifmmode m_t\else $m_t$\fi}
\def\mc{\ifmmode m_c\else $m_c$\fi}
\def\mz{\ifmmode M_Z\else $M_Z$\fi}
\def\mw{\ifmmode M_W\else $M_W$\fi}
\def\mws{\ifmmode M_W^2 \else $M_W^2$\fi}
\def\mhs{\ifmmode m_H^2 \else $m_H^2$\fi}   
\def\mzs{\ifmmode M_Z^2 \else $M_Z^2$\fi}
\def\mts{\ifmmode m_t^2 \else $m_t^2$\fi}
\def\mcs{\ifmmode m_c^2 \else $m_c^2$\fi}
\def\mchs{\ifmmode m_{H^\pm}^2 \else $m_{H^\pm}^2$\fi}
\def\ztwo{\ifmmode Z_2\else $Z_2$\fi}
\def\zone{\ifmmode Z_1\else $Z_1$\fi}
\def\mtwo{\ifmmode M_2\else $M_2$\fi}
\def\mone{\ifmmode M_1\else $M_1$\fi}
\def\tb{\ifmmode \tan\beta \else $\tan\beta$\fi}
\def\xw{\ifmmode x\subw\else $x\subw$\fi}
\def\ch{\ifmmode H^\pm \else $H^\pm$\fi}
\def\lum{\ifmmode {\cal L}\else ${\cal L}$\fi}
\def\inpb{\,{\ifmmode {\mathrm {pb}}^{-1}\else ${\mathrm {pb}}^{-1}$\fi}}
\def\infb{\,{\ifmmode {\mathrm {fb}}^{-1}\else ${\mathrm {fb}}^{-1}$\fi}}
\def\epem{\ifmmode e^+e^-\else $e^+e^-$\fi}
\def\ppb{\ifmmode \bar pp\else $\bar pp$\fi}
\def\bsg{\ifmmode B\to X_s\gamma\else $B\to X_s\gamma$\fi}
\def\bsll{\ifmmode B\to X_s\ell^+\ell^-\else $B\to X_s\ell^+\ell^-$\fi}
\def\bstt{\ifmmode B\to X_s\tau^+\tau^-\else $B\to X_s\tau^+\tau^-$\fi}
\def\lamt{\ifmmode \tilde\lambda\else $\tilde\lambda$\fi}
\def\shat{\ifmmode \hat s\else $\hat s$\fi}
\def\that{\ifmmode \hat t\else $\hat t$\fi}
\def\uhat{\ifmmode \hat u\else $\hat u$\fi}

\newskip\zatskip \zatskip=0pt plus0pt minus0pt
\def\matth{\mathsurround=0pt}
\def\lsim{\mathrel{\mathpalette\atversim<}}

\def\atversim#1#2{\lower0.7ex\vbox{\baselineskip\zatskip\lineskip\zatskip
  \lineskiplimit 0pt\ialign{$\matth#1\hfil##\hfil$\crcr#2\crcr\sim\crcr}}}

\def\grtsim{\,\,\rlap{\raise 3pt\hbox{$>$}}{\lower 3pt\hbox{$\sim$}}\,\,}
\def\lsim{\,\,\rlap{\raise 3pt\hbox{$<$}}{\lower 3pt\hbox{$\sim$}}\,\,}

\renewcommand{\thefootnote}{\fnsymbol{footnote}}

\begin{document} \begin{titlepage}
\rightline{\vbox{\halign{&#\hfil\cr
&SLAC-PUB-12392\cr
}}}
\begin{center}
\thispagestyle{empty} \flushbottom { {
\Large\bf The Determination of the Helicity of $W'$ Boson Couplings at the LHC
\footnote{Work supported in part
by the Department of Energy, Contract DE-AC02-76SF00515}
\footnote{e-mail:
$^a$rizzo@slac.stanford.edu}}}
\medskip
\end{center}

\centerline{Thomas G. Rizzo$^{a}$}
\vspace{8pt} 
\centerline{\it Stanford Linear
Accelerator Center, 2575 Sand Hill Rd., Menlo Park, CA, 94025}

\vspace*{0.3cm}

\begin{abstract}
Apart from its mass and width, the most important property of a new charged gauge boson, $W'$, is the helicity of its 
couplings to the SM fermions. Such particles are expected to exist in many extensions of the Standard Model. In this 
paper we explore the capability of the LHC to determine the $W'$ coupling helicity at low integrated luminosities in the $\ell +E_T^{miss}$ 
discovery channel.  We find that measurements of the transverse mass distribution, reconstructed from this final state in the 
$W-W'$ interference region, provides the best determination of this quantity. To make such measurements requires integrated 
luminosities of $\sim 10(60)~fb^{-1}$ assuming $M_{W'}=1.5(2.5)$ TeV and provided that the $W'$ couplings have Standard Model 
magnitude. This helicity determination can be further strengthened by the use of various discovery channel leptonic asymmetries,  
also measured in the same interference regime, but with higher integrated luminosities. 
\end{abstract}



\renewcommand{\thefootnote}{\arabic{footnote}} \end{titlepage} 

%
%
%

\section{Introduction}

The ATLAS and CMS experiments at the LHC will begin taking data in a few months and it is widely believed that new physics beyond the 
Standard Model(SM) will be discovered in the coming years. There are many expectations as to what this new physics may be and in what 
form it will manifest itself, but it is likely that we will be in for a surprise. Once this new 
physics is discovered our primary goal will be to understand its essential nature and how the specific discoveries, 
such as the production and observed properties of new particles, fit into a broader theoretical framework.

The existence of a new charged gauge boson, $W'$, or a $W'$-like object, is now a relatively common prediction which 
results from many new physics scenarios. These possibilities include the Little Higgs(LH) model{\cite {LH}}, the 
Randall-Sundrum(RS){\cite {RS}} model with bulk gauge fields{\cite {bulk}}, Universal Extra Dimensions(UED){\cite {UED}}, 
TeV scale extra dimensions{\cite {TeV,precision,sphere}},  as well as many different 
extended electroweak gauge models, such as the prototypical Left-Right Symmetric Model(LRM){\cite {LRM,other}}. Although the physics of a new $Z'$ 
has gotten much attention in the literature{\cite {review}}, the detailed study of a possible $W'$ has fared somewhat  
less well{\cite {however}}. Perhaps the most important property of a $W'$, apart from its mass and width, is the helicity of its couplings to the 
fermions in the SM. For all of the models discussed in the literature above, these couplings are either purely left- or 
right-handed, apart from some possible small mixing effects. Determining the helicity of the couplings of a newly discovered $W'$ is thus the 
first major step in opening up the underlying physics as it is an order one discriminator between different classes of 
models.{\footnote {This is similar in nature to determining whether the known light neutrinos are Dirac or Majorana particles.}} 

As will be discussed below, there have been many suggestions over the last 20-plus years as to how to measure the helicity of $W'$ couplings, all of 
which have their own strengths and weaknesses. These analyses have generally relied upon the use of the narrow width approximation. However, 
in employing this approximation much valuable information about the properties of the $W'$ can be lost, in particular, that obtained from 
$W-W'$ interference. 
The goal of this paper will be to explore the effects of this interference on the transverse mass dependent distributions of the $W'$. As we will 
see the rather straightforward measurement of the transverse mass distribution itself will allow us obtain 
the necessary $W'$ helicity information. Furthermore, we will demonstrate that such measurements will require only relatively low integrated luminosities 
for $W'$ masses which are not too large, and will employ the traditional $\ell+E_T^{miss}$ $W'$ discovery channel. 

Section II of the paper contains some background material and a historically-oriented overview of previous ideas that have been suggested to address 
the $W'$ helicity issue including a discussion of their various strengths and weaknesses. Section III will present an analysis of the $W'$ 
transverse mass distribution and its helicity dependence for a range of $W'$ masses, coupling strengths and LHC integrated luminosities. The use of various 
asymmetries evaluated in the $W-W'$ interference region in order to assist with the $W'$ helicity determination will also be discussed.  
Section IV contains a final summary and discussion of our results.

\section{Background and History}

Let us begin by establishing some notation; since much of this should be fairly familiar we will be rather sketchy and refer the interested 
reader to Ref.{\cite {review}} for details. 

We denote the couplings of the SM fermions to the $W_i=(W=W_{SM},W')$ as 
\begin{equation}
\Big({{G_FM_W^2}\over {\sqrt 2}}\Big)^{1/2}V_{ff'}C^{\ell,q}_i\bar f \gamma_\mu (1-h_i\gamma_5)f'W^\mu_i+ h.c.\,,
\end{equation}
where for the case of $W_i=W_{SM}$, the coupling strength(for leptons and quarks, respectively) 
and helicity factors are given by $C^{\ell,q}_i,h_i=1$ and $V_{ff'}$ is the CKM(unit) matrix when 
$f,f'$ are quarks(leptons); note that the helicity structure for both leptons and quarks is assumed to be the same as in all the model cases 
above.{\footnote {For simplicity in what follows we will further assume that the corresponding RH and LH CKM matrices are identical up to phases and 
we will generally neglect any possible small effects arising from $W-W'$ mixing. In the case of RH couplings, we will further assume 
that the SM neutrinos are Dirac fields.}} 
Following the notation given in Ref.{\cite {review}}, with some obvious modifications, the inclusive $pp\to W^+_i \to\ell^+ \nu+X$ differential 
cross section can be written as 
\begin{equation}
{{d\sigma}\over {d\tau~dy~dz}}= K~{{G_F^2M_W^4}\over {48\pi}} \sum_{qq'} |V_{qq'}|^2~\Big[SG_{qq'}^+(1+z^2)+2AG_{qq'}^-z\Big]\,,
\end{equation}
where $K$ is a kinematic/numerical factor that accounts for NLO and NNLO QCD corrections{\cite {nnlo}} as well as leading electroweak
corrections{\cite {electroweak} and is roughly of order $\simeq 1.3$ for suitably defined couplings, $\tau=M^2/s$ ($\sqrt s=14$ TeV at the LHC)  
with $M^2$ being the lepton pair invariant mass. Furthermore,    
\begin{eqnarray}
S &=& \sum_{ij}P_{ij}(C_iC_j)^\ell (C_iC_j)^q(1+h_ih_j)^2\\ \nonumber
A &=& \sum_{ij}P_{ij}(C_iC_j)^\ell (C_iC_j)^q(h_i+h_j)^2\,, 
\end{eqnarray}
where the sums extend over all of the exchanged particles in the $s$-channel. Here  
\begin{equation}
P_{ij}=\shat{{(\shat-M_i^2)(\shat-M_j^2)+\Gamma_i\Gamma_jM_iM_j}\over {[(\shat-M_i^2)^2+\Gamma_i^2M_i^2][i\to j]}}\,,
\end{equation}
with $\shat=M^2$ being the square of the total collision energy and $\Gamma_i$ the total widths of the 
exchanged $W_i$ particles. Note that we have employed $z=\cos \theta$, the scattering angle in the CM frame defined as that between the 
incoming $u$-type quark and the outgoing neutrino (both being fermions 
as opposed to being one fermion and one anti-fermion). Furthermore, the following combinations of parton distribution functions appear: 
\begin{equation}
G^\pm_{qq'}=\Big[q(x_a,M^2)\bar q'(x_b,M^2)\pm q(x_b,M^2)\bar q'(x_a,M^2)]\,,
\end{equation}
where $q(q')$ is a $u(d)-$type quark and $x_{a,b}=\sqrt \tau e^{\pm y}$ are the corresponding parton momentum fractions. 
Analogous expressions can also be written in the case of $W^-_i$ exchange by taking  $z\to -z$ and interchanging initial state quarks and anti-quarks. 

In most cases of interest one usually converts the distribution over $z$ above into one over the transverse mass, $M_T$, formed from the final 
state lepton and the missing transverse energy associated with the neutrino; at fixed $M$, one has $z=(1-M_T^2/M^2)^{1/2}$. The 
resulting transverse mass distribution can then be written as 
\begin{equation}
{{d\sigma}\over {dM_T}} =\int _{M_T^2/s}^1~d\tau \int^Y_{-Y} ~dy~J(z\to M_T)~{{d\sigma}\over {d\tau~dy~dz}}\,,
\end{equation}
where $Y=min(y_{cut},-1/2 \log \tau)$ allows for a rapidity cut on the outgoing leptons and $J(z\to M_T)$ is the appropriate 
Jacobian factor{\cite {BP}}. In practice, $y_{cut}\simeq 2.5$ for the two LHC detectors.   
Note that ${{d\sigma}\over {dM_T}}$ will only pick out the $z$-even part of ${{d\sigma}\over {d\tau~dy~dz}}$ as well as the even combination of 
terms in the product of the parton densities, $G^+_{qq'}$. 
In the usual analogous fashion to the $Z'$ case{\cite {review}}, as we will see in our discussion below, 
one can also define the forward-backward asymmetry as a function of the transverse mass, in principle prior to integration over the rapidity $y$,  
$A_{FB}(M_T,y)$, whose numerator now picks out the $z$-odd terms in  ${{d\sigma}\over {d\tau~dy~dz}}$ as well as the odd combination of 
terms in the parton densities $G^-_{qq'}$. 

To be complete, we note that 
historically when discussing new gauge boson production, particularly when dealing with states which are weakly coupled as will be the case 
in what follows, use is often made of the narrow width approximation(NWA). In the $W'$ case of relevance here, the NWA essentially replaces 
the integration over $d\tau \sim dM$ by a $\delta$ function, \ie, the $W'$ is assumed to be produced on-shell. Thus, 
for any smooth function $f(M)$, essentially, $\int ~dM~f(M)$ 
$\to \int ~dM~f(M)~{{\pi}\over {2}}\Gamma_{W'}\delta (M-M_{W'})$ $\to {{\pi}\over {2}}\Gamma_{W'}f(M_{W'})$, apart from some overall factors. 
Note that use of the NWA implies that we evaluate quantities on the `peak' of the $W'$ mass distribution, \ie , at $M=M_{W'}$. 
This approximation is usually claimed to be valid up to $O(\Gamma_{W'}/M_{W'})$ corrections(at worst), 
but there are occasions, \eg, when $W-W'$ interference is important, when its use can lead to a loss of 
valuable information and may even lead to wrong conclusions{\cite {NWA}}. 
Unfortunately, in the $W'$ case,  the quantity $M$ itself is not a true observable due to the missing longitudinal momentum of the neutrino. 

Given this background, let us now turn to an historical discussion of the determination of the $W'$ coupling helicity. To be concrete, we will consider 
two different $W'$ models; we will assume for simplicity that $C^{\ell,q}_{W'}=1$ in both cases and that only the value of $h_{W'}=\pm 1$ 
distinguishes them. In this situation, employing the NWA, the cross section for on-shell $W'$ production (followed by its leptonic decay)  
is proportional to $\sim (1+h_{W'}^2)$ and is trivially seen to be independent of 
the helicity of the couplings. We would thus conclude that cross section measurements are not useful helicity discriminants. 
More interestingly, as was noted long ago{\cite {Haber:1984gd}}, we find that the rapidity integrated value of $A_{FB}$, given in 
the NWA by 
\begin{equation}
A_{FB} \sim {{h_{W'}^2}\over {(1+h_{W'}^2)^2}}\,,
\end{equation}
also has the {\it same value} for either purely LH or RH couplings{\footnote {This follows immediately from the fact that we have assumed that 
both the hadronic and leptonic couplings of the $W'$ have to have the {\it same} helicity.}}. Thus, in the NWA, $A_{FB}$ provides no help 
in determining the $W'$ coupling helicity structure for the cases we consider here. However, we note that if the quark and leptonic coupling 
helicities of the $W'$ are {\it opposite}, then the value of $A_{FB}$ will flip sign in comparison to the above expectation.  

It is apparent from this result that some other observable(s) must be used to distinguish these 
two cases. Keeping the NWA assumption, the first suggestion{\cite {Haber:1984nh}} along these 
lines was to examine the polarization of $\tau$'s originating in the 
decay $W'\to \tau \nu$. In that paper it was explicitly shown that the the energy spectrum of the final state particle in the decay 
$\tau \to \ell, \pi$ or $\rho$ (in the $\tau$ rest frame) 
was reasonably sensitive to the original $W'$ helicity since the $\tau$ itself effectively decays only 
through the SM LH couplings of the $W$(provided the $W'$ is sufficiently massive as we will assume here). The difficulty with this method is that the 
observation of this decay mode at the LHC is not all that straightforward and even the corresponding $Z'\to \tau \tau$ mode, which is somewhat easier 
to observe, is just beginning to be studied by the LHC experimental collaborations{\cite {Vickey}}. Clearly, measuring the polarization of the 
$\tau$'s in $W'\to \tau \nu$ will be reasonably difficult in the LHC detector environment and 
may, at the very least, require large integrated luminosities even for a relatively light $W'$. The results of detailed studies by the LHC collaborations 
to address this issue are anxiously awaited.  

In the early 90's, two important NWA-based methods for probing the helicity of the $W'$ were suggested{\cite {Cvetic}}. The first of these  
is an examination of the rare decay mode $W'\to \ell^+\ell^-W$ (with the $W$ decaying into jets); in particular, one makes a measurement of 
the ratio of branching fractions  
\begin{equation}
R_W={{B(W'\to \ell^+\ell^-W)}\over {B(W'\to \ell \nu)}}\,,
\end{equation}
obtained by employing the NWA. $R_W$ is expected to be roughly $\sim$O(0.01) or so after suitable cuts. 
One of the main SM backgrounds, \ie, $WZ$ production, can essentially 
be removed by demanding that the dileptons do not form a $Z$, demanding that the mass of the $jj\ell \ell$ 
system be not far from the (already known) value of $M_{W'}$ and that of the dijets reconstructs to the $W$ mass. Even after there requirements, however, 
some background from the 
continuum would remain. Furthermore, as the energy of the final state $W$ increases it is more likely that the resulting dijets will coalesce 
into a single jet depending on the jet cone definition which is employed. In this case, at the very least, a very large additional background from  
single jets may appear; it is also possible that the events with a final state $W$ would be completely lost without the dijet mass reconstruction. 
The $3\ell+E_T^{miss}$ final state, with suitable cuts, would be obviously cleaner and would avoid some of these issues but at the price of an overall 
suppression due to ratio of branching fractions of $\simeq 1/3$ thus reducing the mass range over which this process would be useful.

In a general gauge model, the amplitude for this process is the sum of two graphs. In the first graph, 
$W^{'-}\to \ell^- \bar \nu^*$, \ie, the production of a virtual neutrino followed by the `decay' $\bar \nu^* \to \ell^+ W^-$. Clearly, if the $W'$ couples 
in a purely RH manner to the SM leptons then this graph will vanish in the limit of massless neutrinos due to the presence of two 
opposite helicity projection 
operators. This graph will, of course, be non-zero only if the $W'$ couples in an at least partially LH manner. The second graph involves 
the presence  
of the trilinear couplings $W'ZW$ and $W'Z'W$; recall that in any model with a $W'$, a $Z'$ will also appear just based on gauge invariance. 
In this case, the decay proceeds as $W'\to WZ/Z^{'*} \to W\ell^+\ell^-$, noting that the on-shell SM 
$Z$ contribution can be removed by a suitable cut on 
the dilepton invariant mass. The main issue is the size of the $W'Z'W$ (and $W'ZW$) couplings and this can involve such things such as, \eg, 
the detailed electroweak symmetry breaking patterns of the given model under study. {\it Generically} in extra dimensional models{\cite {bulk,
UED,TeV,precision,sphere}}, these couplings are absent in the limit of small mixing due the orthogonality of the Kaluza-Klein wavefunctions 
of the states. In models where the SM $SU(2)_L$ arises from a diagonal breaking of the form $G_1 \otimes G_2 \to SU(2)_{Diag}$, such as in LH 
models{\cite {LH}}, the $W'Z'W$ coupling is of order the SM weak coupling, $g$, while the  $W'ZW$ coupling is either of order $g$ or can be 
mixing angle suppressed. In other cases, such as in  
the LRM{\cite {LRM}}, where $SU(2)_L\otimes SU(2)_R$ just breaks to $SU(2)_L$, the  $W'ZW,WZ'W$ couplings are only generated by mixings and for the 
diagrams of interest are not longitudinally enhanced. Since the amplitude associated with the pure leptonic graphs are absent in this case, the 
entire amplitude is mixing angle suppressed so that this process has an unobservably small rate. In fact, there are no known models where the 
$W'$ helicity is RH and the $W'ZW,WZ'W$ couplings are not mixing angle suppressed{\footnote {In a {\it fundamental} UV complete theory, this may follow 
directly from arguments based solely on gauge invariance and the requirement of high energy unitarity.}}. Thus, {\it based on known models}, 
it appears that the observation of the rare decay $W'\to \ell^+\ell^-W$ would be a compelling indication that the $W'$ is at least partially coupled 
in a LH manner 
with apparently most of the serious SM backgrounds being 
removable by conventional cuts. However, in making a truly model-independent analysis one must exercise care in the use of this result. A detailed 
analysis of the signal and backgrounds, including that for the $jj\ell^+\ell^-$ final state, for such decays including realistic detector effects would 
be very useful in addressing all these issues and should be performed. However, it also seems clear that is unlikely that a reliable measurement of 
$R_W$ can be made with relatively low integrated luminosities.  

A second, imaginative possibility is to observe $WW'$ associated production{\cite {Cvetic}} with $W\to jj$ for the same reasons as above. 
Many of the arguments made in the 
previous paragraph will also apply in this case as well since the diagrams responsible for this process are quite similar to previously discussed. 
Essentially these graphs are obtained by crossing, with the final state leptons now replaced by an initial state $q\bar q$. In this case one looks 
for the $jj\ell E_T^{miss}$ final state with the $\ell E_T^{miss}$ transverse mass peaking near $M_{W'}$. One would anticipate this cross section to 
be of order $\sim 0.01$ of that of the $W'$ discovery channel. The main issues here are, as above, the SM backgrounds and the nature 
of the triple gauge vertices. It is not likely that a reliable measurement of this cross section will be performed with low luminosities that 
could be interpreted in a model-independent way until all of the background and detector issues are dealt with. Again, a detailed analysis including 
detector effects should be performed.

\begin{figure}[htbp]
\centerline{
\includegraphics[width=7.5cm,angle=90]{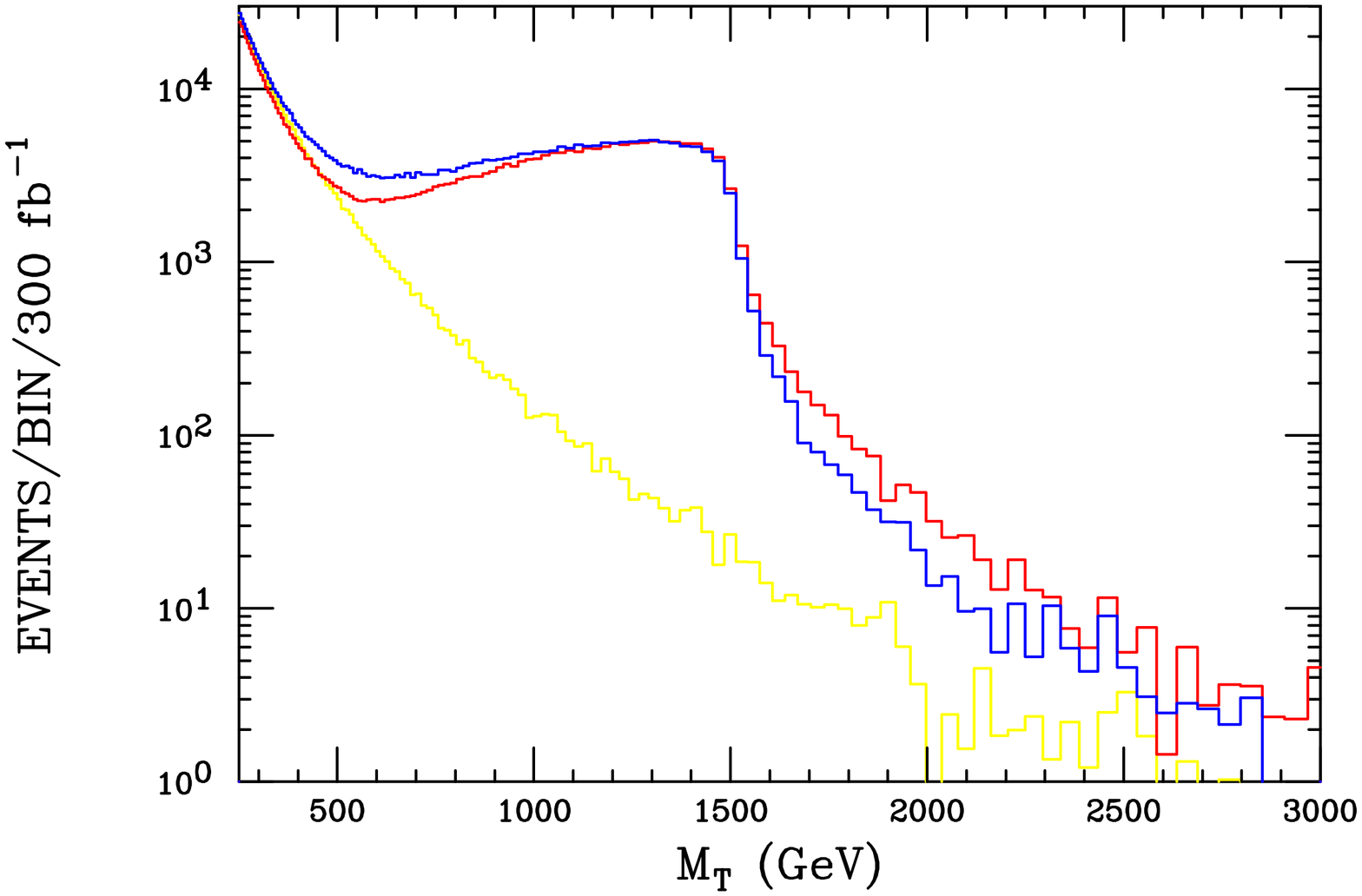}}
\vspace*{0.1cm}
\centerline{
\includegraphics[width=7.5cm,angle=90]{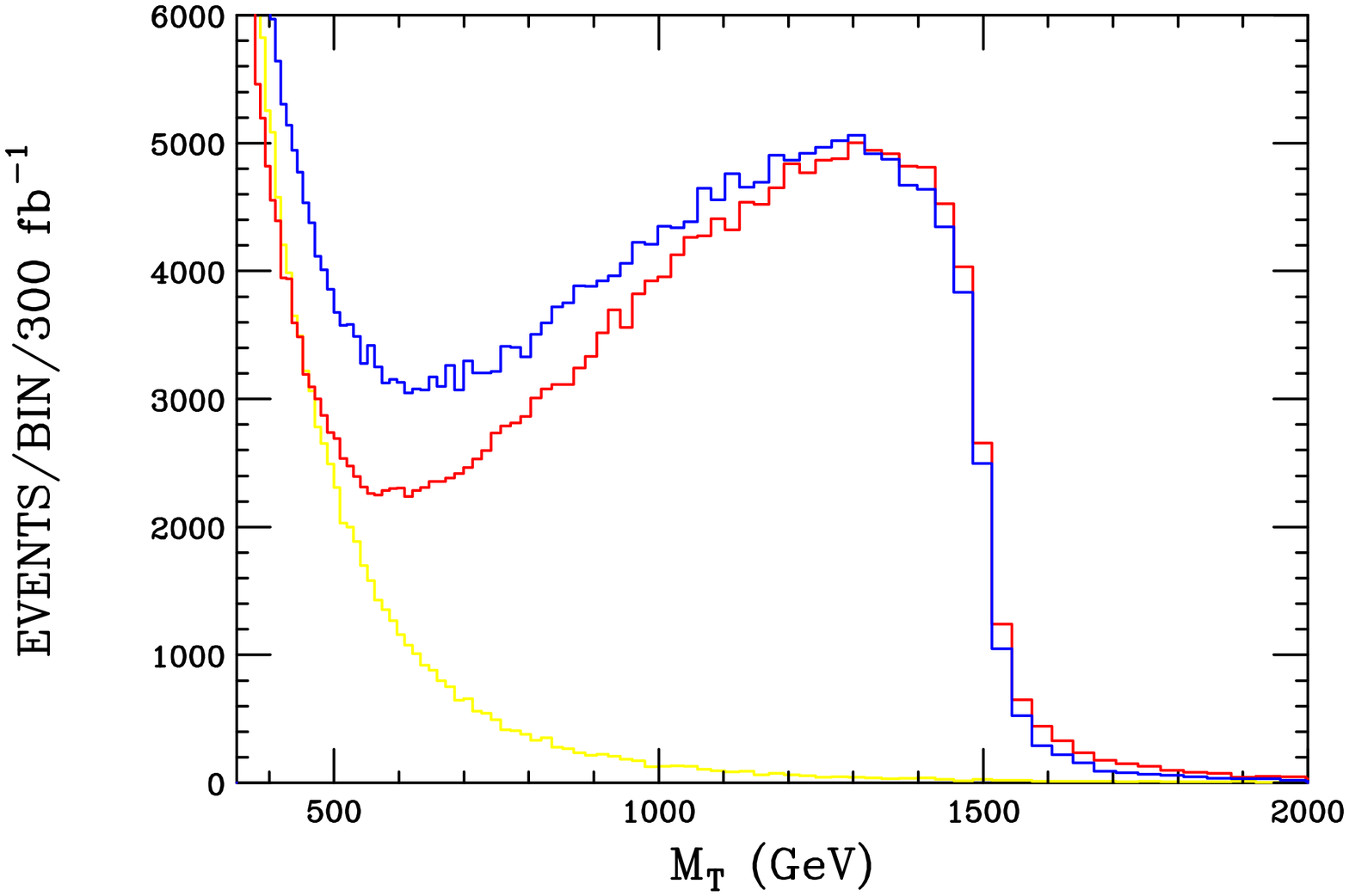}}
\vspace*{0.1cm}
\caption{Transverse mass distribution for the production of a 1.5 TeV $W'$ including interference effects at the LHC displayed on both log and 
linear scales assuming an integrated luminosity of $300~fb^{-1}$. The lowest histogram is the SM continuum background. The upper blue(middle red) 
histogram at $M_T=600$ GeV corresponds to the case of $h_{W'}=-1(1)$.}
\label{fig1}
\end{figure}

\section{$W-W'$ Interference as a Function of $M_T$}

What we have learned from the previous discussion is that tools which employ the NWA are 
not particularly useful when we are trying to determine the $W'$ coupling 
helicity with relatively low luminosities in an easily examined final state. One of the key reasons for this is that the use of NWA does not 
allow us to examine the influence of $W-W'$ interference to which we now turn{\cite {Boos:2006xe}}{\footnote {We note in passing that the usual 
experimental analyses at LHC{\cite {analyses}} performed by both the ATLAS and CMS collaborations (as well as those at the Tevatron by CDF and D0{\cite {limit}}) 
ignore the effects of $W-W'$ interference since these contributions are absent from 
default versions of stand-alone PYTHIA{\cite {pythia}}.}}. To be specific, in the analysis that follows, we will 
employ the CTEQ6M parton densities{\cite {CTEQ}} and will restrict our attention only to the $\ell=e$ final state since it is better measured at 
these energies{\cite {analyses}} yielding a better $M_T$ resolution. Furthermore, we will 
assume that only SM particles are accessible in the decay of the $W'$ so that the total width can be straightforwardly calculated from the assumptions 
described above and its assumed mass value; for example, we obtain $\Gamma(W')=51.9$ GeV assuming a $W'$ mass of 1.5 TeV including QCD corrections. 
NLO QCD modifications to the distributions we discuss below have been ignored but those distributions we consider are rather 
robust against large corrections. 

\begin{figure}[htbp]
\centerline{
\includegraphics[width=7.5cm,angle=90]{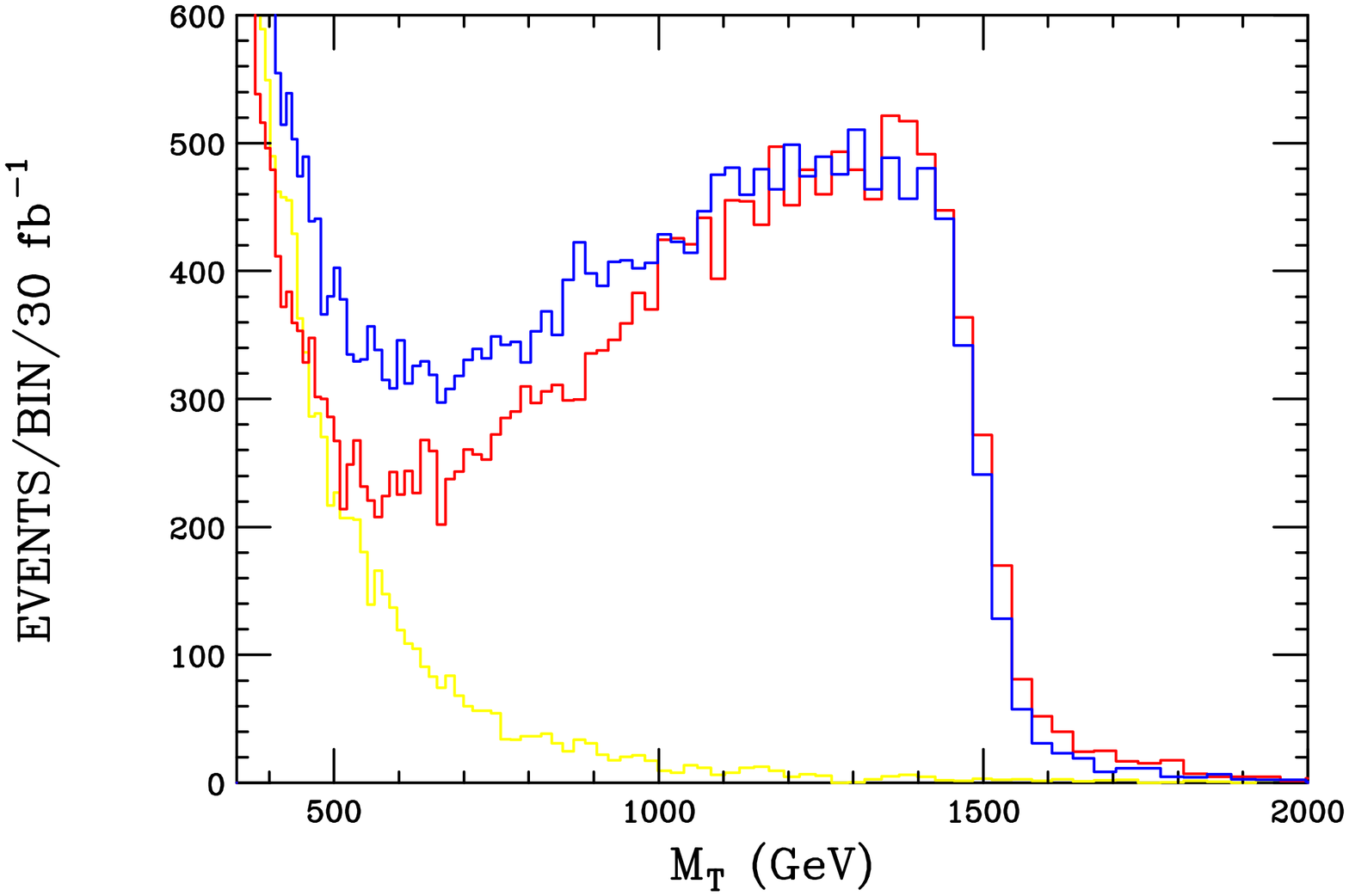}}
\vspace*{0.1cm}
\centerline{
\includegraphics[width=7.5cm,angle=90]{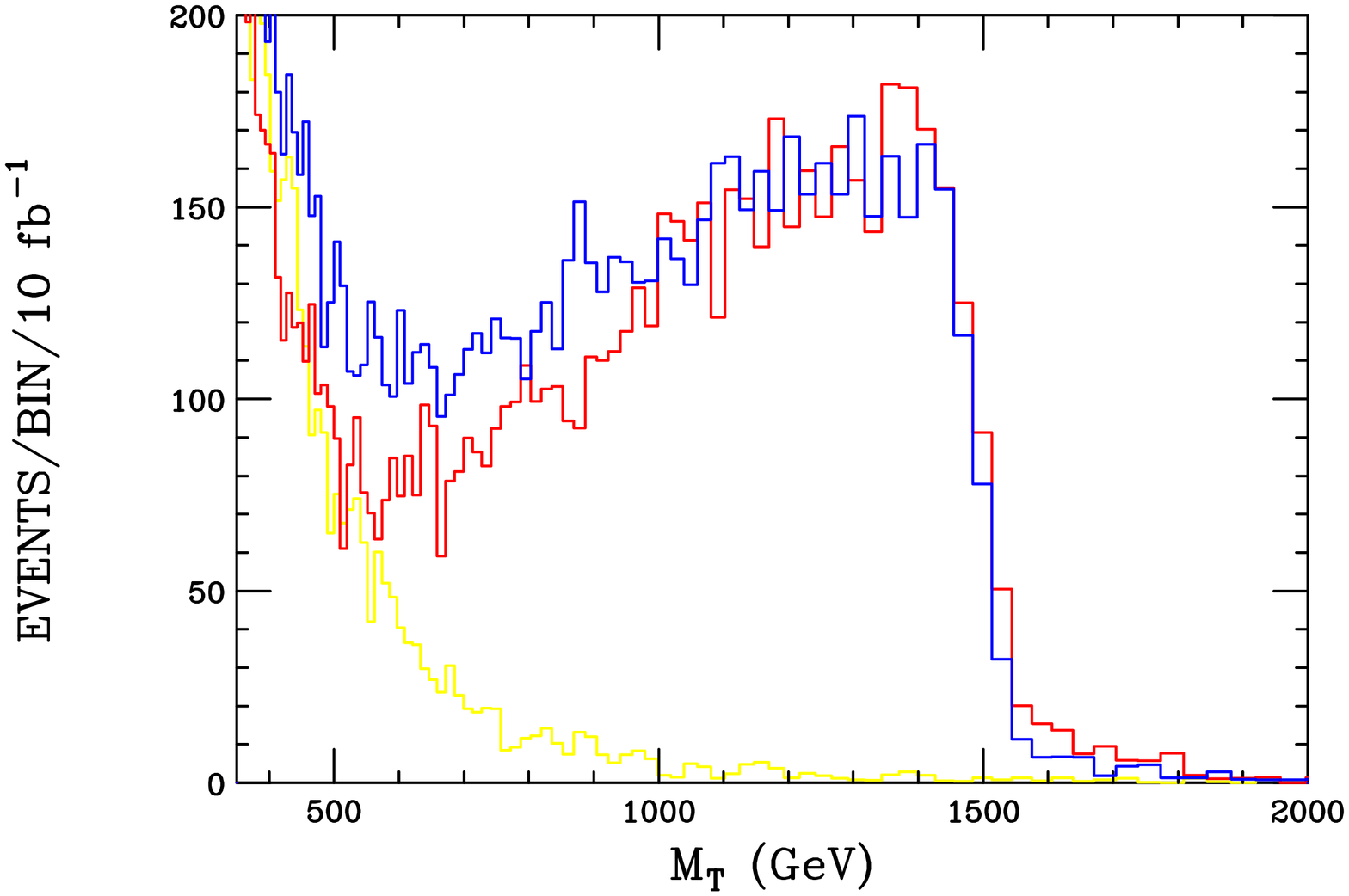}}
\vspace*{0.1cm}
\caption{Same as in the previous figure but now on a linear scale with lower luminosities and smeared by the detector resolution. In the top(bottom) 
panel an integrated luminosity of 30(10) $fb^{-1}$ has been assumed. Detector smearing has now been included assuming $\delta M_T/M_T=2\%$.}
\label{fig2}
\end{figure}

The most obvious distribution to examine first is ${{d\sigma}\over {dM_T}}$ itself; for the moment let us restrict ourselves to the two cases where 
$C^{\ell,q}_{W'}=1$ and $h_{W'}=\pm 1$. Fig.~\ref{fig1} shows this distribution for a large integrated luminosity, assuming 
$M_{W'}=1.5$ TeV{\cite {limit}}, as well as the SM continuum background{\footnote {Note that we would expect to see 
many excess events for such $W'$ masses as only $\simeq 25~pb^{-1}$ of luminosity would be needed to discover($5\sigma$) such as state at the LHC.}}.  
In obtaining these and other $M_T$-deprndent distributions below, a cut on the lepton rapidity, $|\eta_\ell \leq 2.5$, has been applied. 
Several things are immediately clear: 
($i$) In the region near the Jacobian peak both distributions are quite similar; this is not surprising as this is the region where the NWA is 
most applicable since now $M_T \simeq M$ and $W-W'$ interference is minimal. 
In this limit we would indeed recover our earlier result that the cross section is helicity independent. 
($ii$) In the lower $M_T$ region where interference effects are important the two models lead to quite different distributions. In particular, 
for the LH case with $h_{W'}=1$, 
we observe a destructive interference with the SM amplitude producing a distribution that lies below that of the pure SM continuum background. 
(This is not surprising as the overall signs of the $W$ and $W'$ contributions are the same but we are at values of $\sqrt {\shat}$ that are 
above $M_W$ yet below $M_{W'}$ so that the relevant propagators have opposite signs.) However, for the RH case with $h_{W'}=-1$, there is no such 
interference and therefore the resulting distribution always lies above the SM background. It is fairly obvious that these two distributions 
are trivially distinguishable at these large integrated luminosities. 
Note that other contributions to the SM background, \eg, those from the decay of top quarks as well as guage boson pairs, have been shown to be rather 
small at these masses at the detector level {\cite {analyses}}, at the level of a few percent, and will be ignored in the analysis that follows.  

\begin{figure}[htbp]
\centerline{
\includegraphics[width=7.5cm,angle=90]{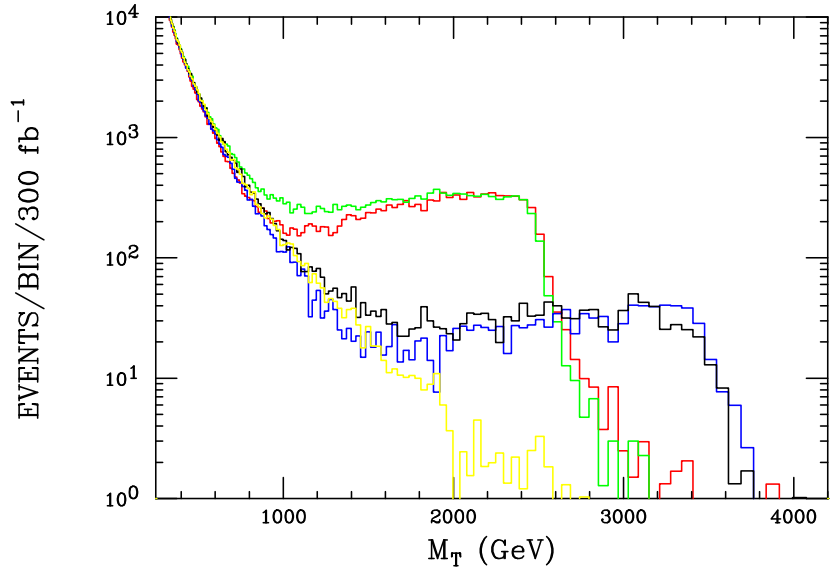}}
\vspace*{0.1cm}
\caption{High luminosity plot of the transverse mass distribution assuming $M_{W'}=2.5(3.5)$ for the upper(lower) pair of histograms along with the 
SM continuum background. In the interference region near $\simeq 0.5M_{W'}$ the upper(lower) member of the pair corresponds to the case of 
$h_{W'}=-1(1)$. Detector smearing has now been included assuming $\delta M_T/M_T=2\%$. }
\label{fig3}
\end{figure}

Fig.~\ref{fig2} shows the same ${{d\sigma}\over {dM_T}}$ distribution on a linear scale 
but now for far smaller integrated luminosities that may be obtained during early LHC running; here we include the effects of detector 
smearing, with $\delta M_T/M_T \simeq 2\%$, which is somewhat less important in the very large statistics sample cases shown above.  
It is immediately apparent that even with only $\sim 10~fb^{-1}$ of luminosity 
the two cases remain quite distinct; however, it also appears unlikely that much smaller luminosities would be very useful in this regard. This result 
is a significant improvement over previous attempts to determine the $W'$ coupling helicity with low luminosities in clean channels. 

\begin{figure}[htbp]
\centerline{
\includegraphics[width=7.5cm,angle=90]{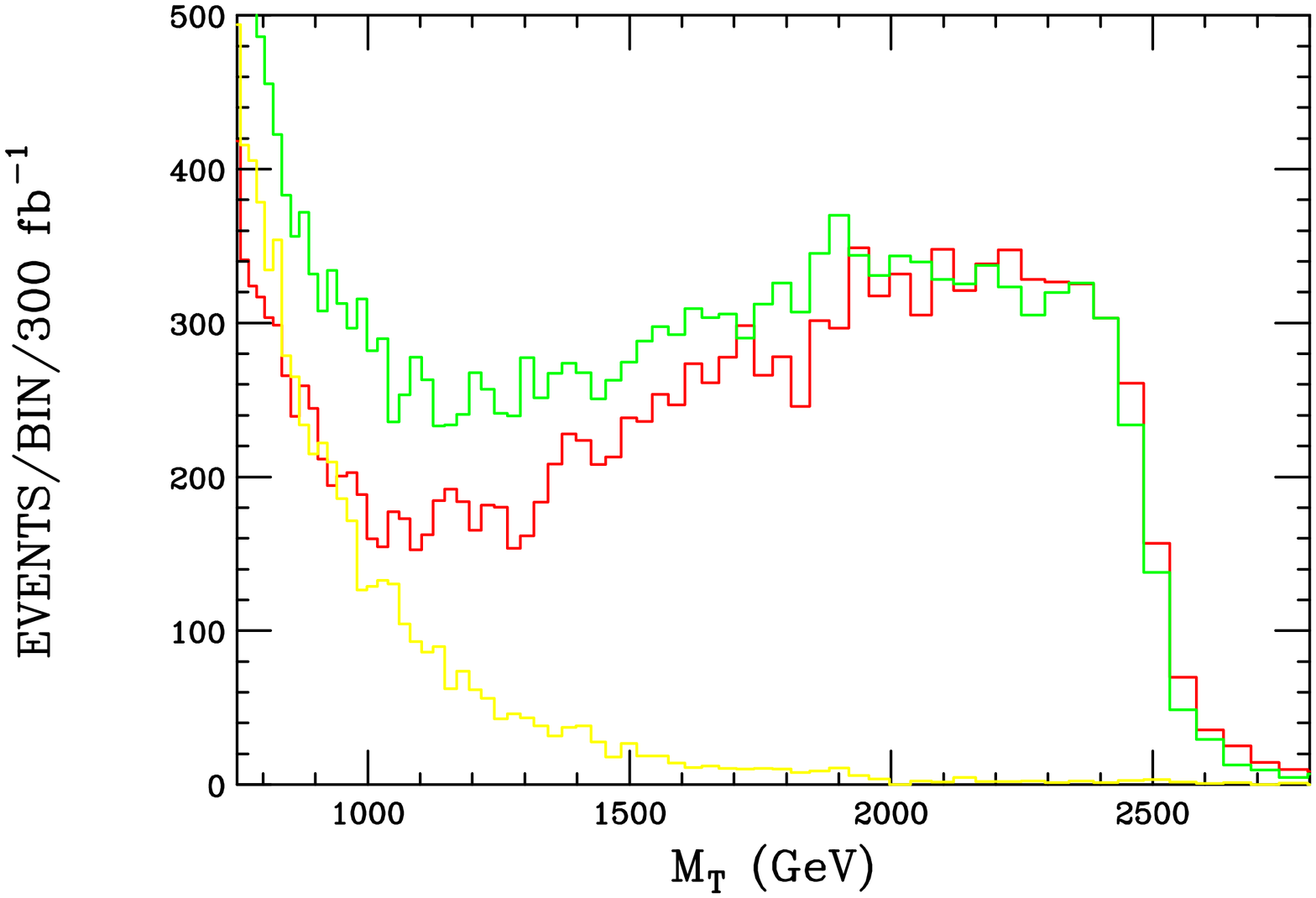}}
\vspace*{0.1cm}
\centerline{
\includegraphics[width=7.5cm,angle=90]{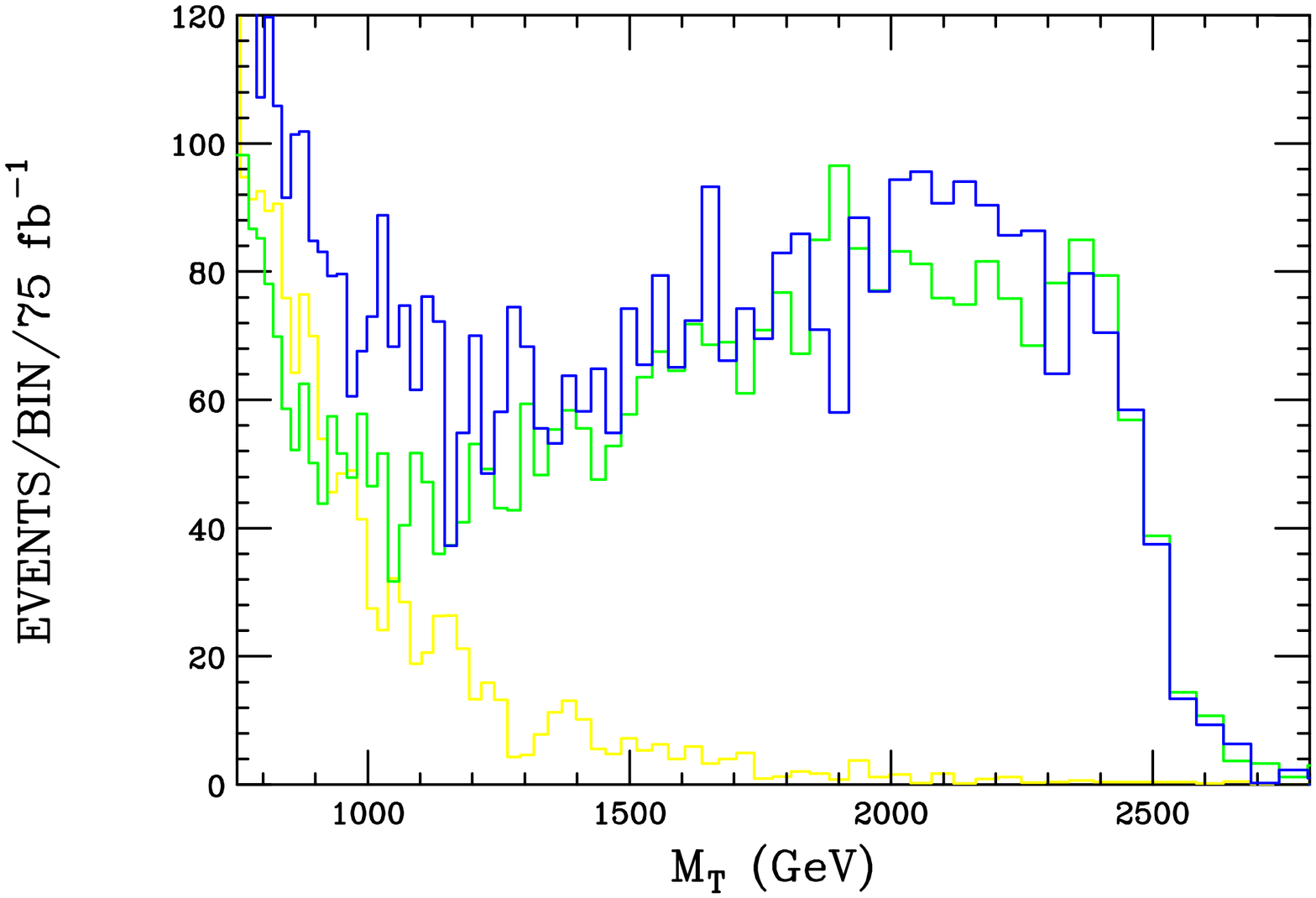}}
\vspace*{0.1cm}
\caption{Transverse mass distribution assuming a mass of 2.5 TeV for the $W'$ along with the SM continuum background; the upper(lower) panel 
corresponds to a luminosity of 300(75) $fb^{-1}$. In the interference region near $\simeq 0.5M_{W'}$ the upper(lower) histogram corresponds to 
the case of $h_{W'}=-1(1)$.}
\label{fig4}
\end{figure}

At this point there are several important questions one might ask: ($i$) What happens for a more massive $W'$, \ie, how much luminosity will be 
needed in such cases to distinguish $W'$ couplings of opposite helicities? ($ii$) What if the the $W'$ couplings are weaker than our canonical 
choice above? 
($iii$) Do other observables, \eg, $A_{FB}$, measured in the interference region below the Jacobian peak assist us in model separation? 
($iv$) In the case where the $W'$ is a KK excitation, does the presents of the additional $W$ KK tower members alter these results? 
($v$) In the discussion above we have assumed that $C_{W'}^\ell=C_{W'}^q$; what would happen, \eg, if their signs were opposite thus modifying the 
interference bewteen the $W$ and $W'$? 
($vi$) What if the $W'$ couplings are not purely chiral and are an admixture of LH and RH helicities? It is to these issues that we now turn.
  
Fig.~\ref{fig3} provides us with a high luminosity overview for the more massive cases where $M_{W'}=2.5$ or 3.5 TeV. In the  $M_{W'}=2.5$ TeV case, 
Fig.~\ref{fig4} demonstrates that the full 300 $fb^{-1}$ luminosity is not required to distinguish the two possibly helicities; $\sim 60 fb^{-1}$ 
seems to be the approximate minimum luminosity that appears to be 
necessary. For higher masses, distinguishing the two cases becomes far more difficult due to the smaller production cross section as we see from 
Fig.~\ref{fig5} for the case of $M_{W'}=3.5$ TeV assuming a luminosity of 300 $fb^{-1}$; essentially the full luminosity is required for model 
distinction in this case. 

\begin{figure}[htbp]
\centerline{
\includegraphics[width=7.5cm,angle=90]{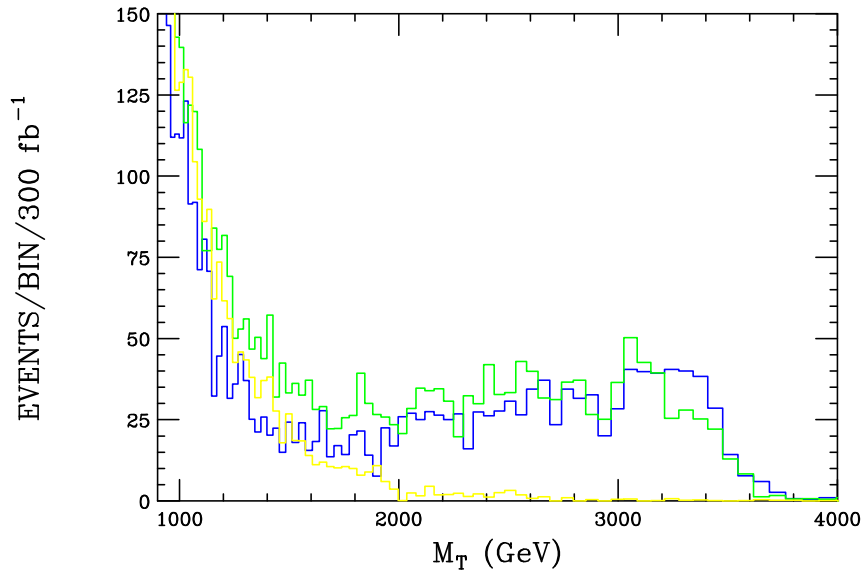}}
\vspace*{0.1cm}
\caption{High luminosity plot of the transverse mass distribution assuming $M_{W'}=3.5$ TeV along with the SM continuum background. 
In the interference region near $\simeq 0.5M_{W'}$ the upper(lower) histogram corresponds to the case of $h_{W'}=-1(1)$. }
\label{fig5}
\end{figure}

What if the $W'$ couplings are weaker? Clearly if they are too weak there will be insufficient statistics to discriminate the two 
possible coupling helicity assignments for any fixed value of $M_{W'}$. In order to examine a realistic example of this situation, 
we consider the case of the second $W$ KK excitation in the 
UED model{\cite {UED,narrow}} with a conserved KK-parity. In such a scenario the LH couplings of this field to SM fermions vanish at tree level but are induced 
by one loop effects. In this case one finds that the effective values of $C^{\ell,q}$ are distinct but are qualitatively of order $\sim 0.05$ though we 
employ the specific values obtained in Ref.{\cite {UED,narrow} below in the actual calculations. 
Fig.~\ref{fig5p} shows the transverse mass distributions in this case assuming that $M_{W'}=$1 TeV for the second level KK state. The signal for this 
$W$ KK state is clearly visible above the SM background. However, we also see that for even for these high luminosities and low masses 
the two helicity choices are not distinguishable. Clearly, one cannot determine the $W'$ coupling helicity for such very weak interaction strengths. 
Semi-quantitatively, 
we find that that this breakdown in the discriminating power occurs when  $(C^\ell C^q)^{1/2}\sim 0.1$ at these luminosities and masses. 

\begin{figure}[htbp]
\centerline{
\includegraphics[width=7.5cm,angle=90]{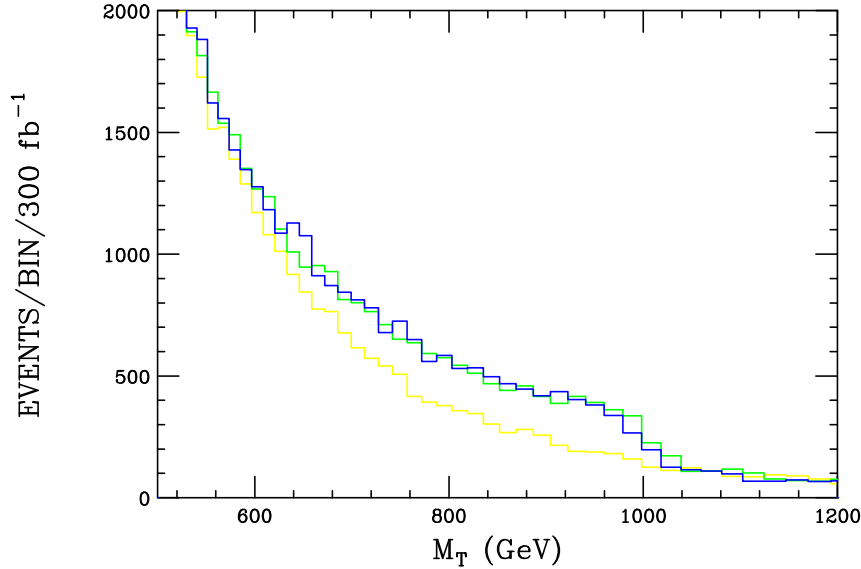}}
\vspace*{0.1cm}
\caption{High luminosity plot of the transverse mass distribution assuming $M_{W'}=1$ TeV for the second $W$ KK level in the UED model 
smeared by detector resolution as above. As usual the lower histogram is the SM background while the other two correspond to the signal cases 
with $h_{W'}=-1(1)$ and are essentially indistinguishable.}
\label{fig5p}
\end{figure}

We now turn to the next question we need to address: 
can asymmetries be useful in strengthening our ability to determine the $W'$ coupling helicity? We know from the discussion above that the answer is 
apparently `no' in the NWA limit, \ie, when $M_T \simeq M$. Thus we must focus our attention on the $M_T$ region below the peak where $W-W'$ interference 
is strongest or, more generally, examine the asymmetries' $M_T$-dependence directly. 
The most obvious quantity to begin with is the $y$-integrated value of $A_{FB}$ 
for both $W'^{\pm }$ channels.  To make such a measurement, we need to know several things in addition to the sign of the lepton (which we 
assume can be done with $\simeq 100\%$ efficiency). At the parton level, in the case of $W'^{-}$ for example, the relevant angle used to define 
$A_{FB}$ lies between the incoming $d$-type quark and the outgoing $\ell^-$. Reconstructing this direction presents us with two problems: first, 
since the longitudinal momentum of the $\nu$ is unknown there is an, in principle, two-fold ambiguity in the motion of the center of mass in the 
lab frame; this can cause a serious dilution of the observed asymmetry but can be corrected for statistically using Monte Carlo once the $W'$ mass is known. 
Second, even when it is determined, the direction of motion of the center of mass is not necessarily that of the $d$-type quark though it is likely 
to be so when the boost of the center of mass frame is large. The later problem also arises 
for the case of a $Z'$ and has also been shown to be mostly correctable in detailed Monte Carlo studies{\cite {afb}}. For the moment, let us forget these 
issues and ask what the $y$-integrated $A_{FB}(M_T)$ looks like in both $\ell^\pm$ channels; the results are shown in Fig.~\ref {fig6} assuming 
high luminosities and $M_{W'}=1.5$ TeV. Here we see that these integrated quantities, even for luminosities of 300 $fb^{-1}$, are essentially useless 
in distinguishing the two coupling helicity cases. Furthermore, we also see that the two coupling helicities lead to essentially identical results when 
$M_T \simeq M_{W'}$ as would be expected based on the NWA. A short analysis indicates that approximately ten times more integrated luminosity would be 
required before some separation in the two cases becomes possible{\cite {slhc}}. Clearly this situation would only become worse if we were to raise the mass 
of the $W'$ or reduce its coupling strength.

\begin{figure}[htbp]
\centerline{
\includegraphics[width=7.5cm,angle=90]{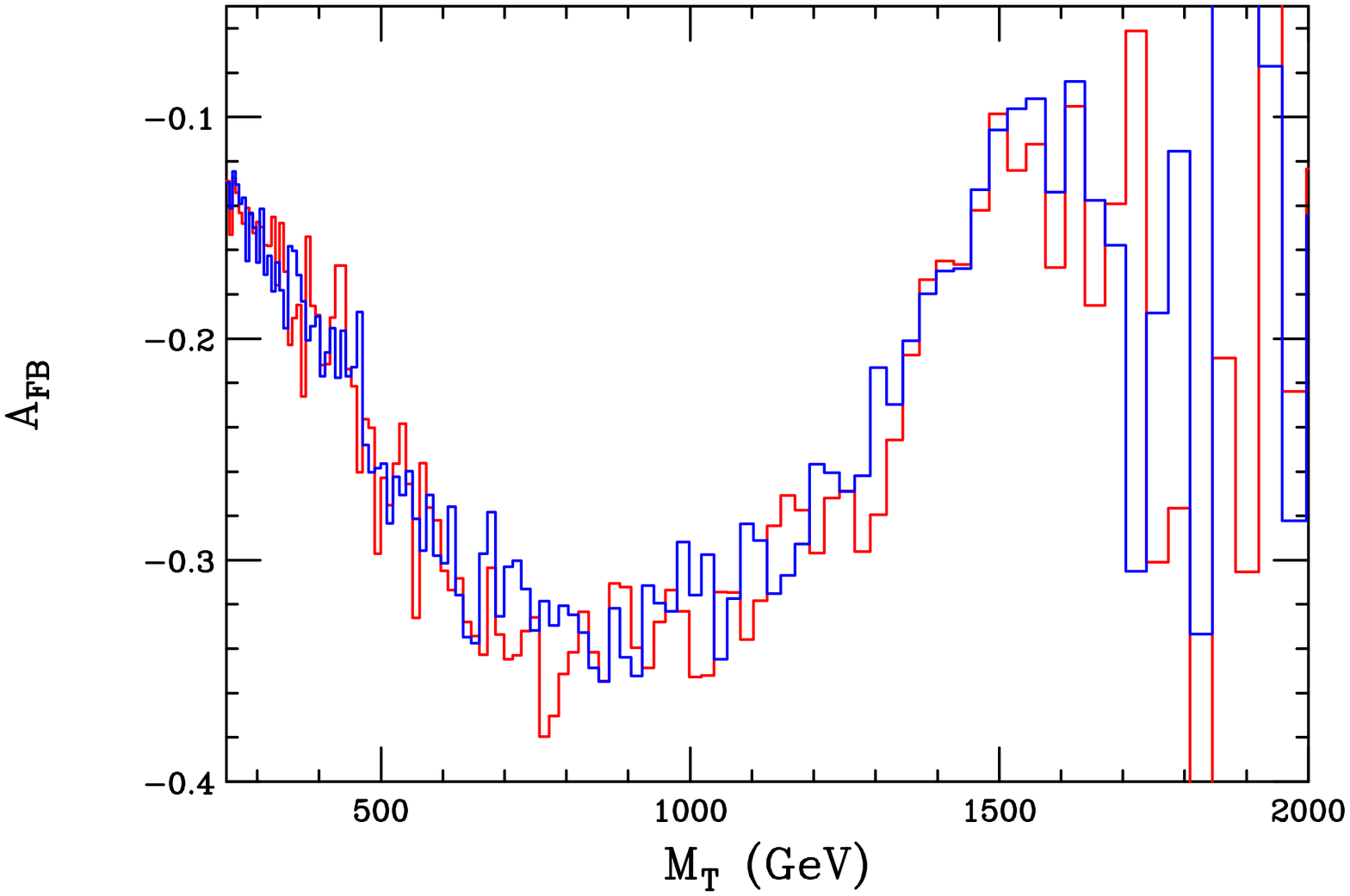}}
\vspace*{0.1cm}
\centerline{
\includegraphics[width=7.5cm,angle=90]{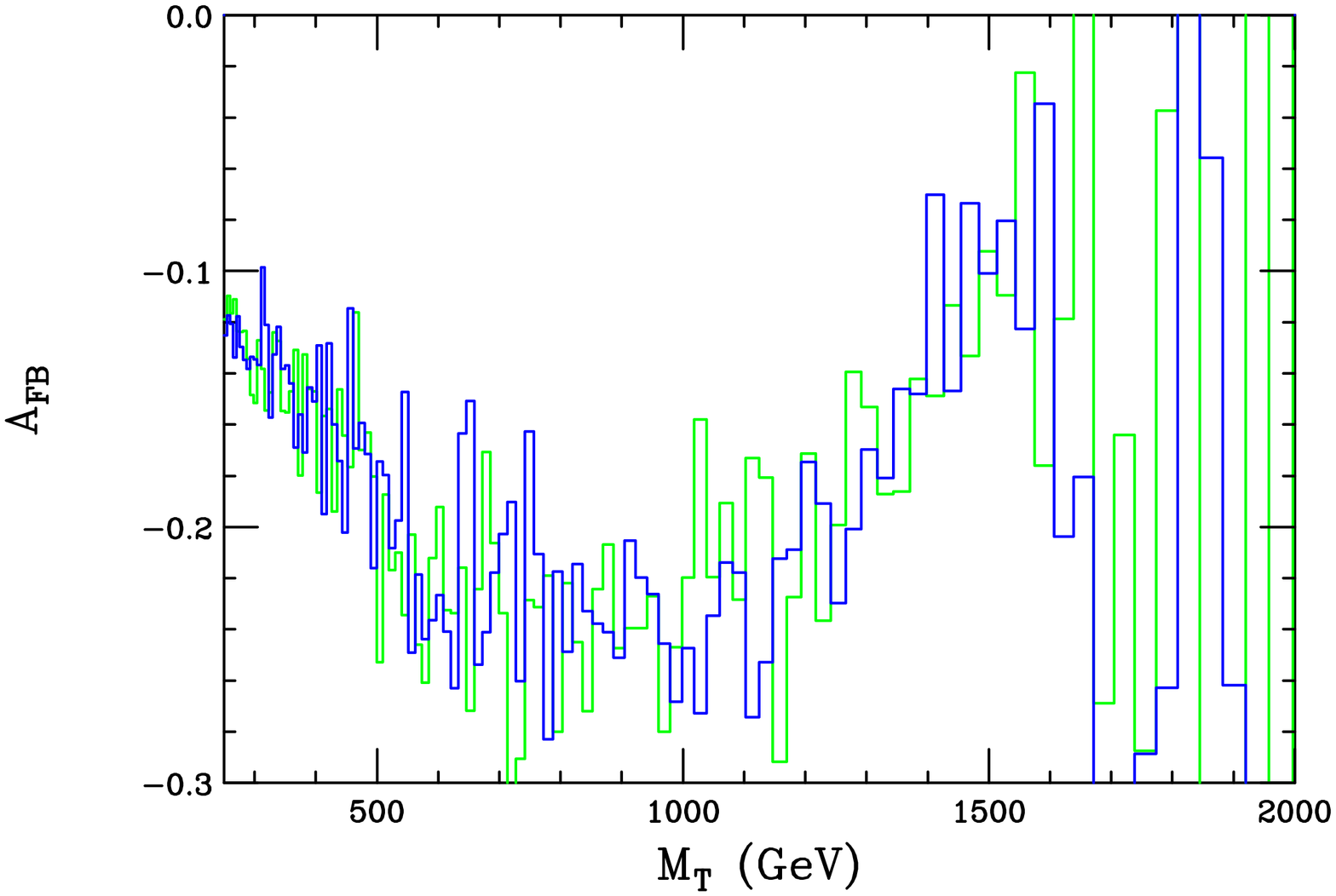}}
\vspace*{0.1cm}
\caption{The $y$-integrated value of $A_{FB}$, as a function of the transverse mass, assuming a mass of 1.5 TeV for the $W'^+$($W'^-$) in the 
top(bottom) panel. Here an integrated luminosity of 300 $fb^{-1}$ has been assumed. The two essentially indistinguishable histograms correspond to the two 
possible choices of the helicity, $h_{W'}=\pm 1$.}
\label{fig6}
\end{figure}

It is perhaps possible that some information is lost by only using the integrated quantity $A_{FB}$ and we need to consider instead $A_{FB}(y_W)$, where 
$y_W$ is the rapidity of the center of mass frame. This distribution is odd under the interchange $y_W \to -y_W$ at the LHC so we can simply fold 
this distribution over the $y_W=0$ boundary to double the statistics. Furthermore, by integrating over a wide $M_T$ range in the interference region 
below the $W'$ peak, \eg, $0.4 \leq M_T \leq 1$ TeV in the case of a 1.5 TeV $W'$, further statistics can be 
gained. Fig~\ref{fig7} shows the resulting $A_{FB}(y_W)$ distributions for a $W'^{\pm}$ with mass of 
1.5 TeV assuming a luminosity of 300 $fb^{-1}$ for $h_{W'}=\pm 1$. At these large luminosities, the $A_{FB}(y_W)$ distributions for the two helicity 
choices are clearly distinguishable but this will certainly become more difficult for lower luminosities or for larger masses. We find that we essentially 
loose all coupling helicity information when the luminosity falls much below $\simeq 100 fb^{-1}$ for this $W'$ mass.

\begin{figure}[htbp]
\centerline{
\includegraphics[width=7.5cm,angle=90]{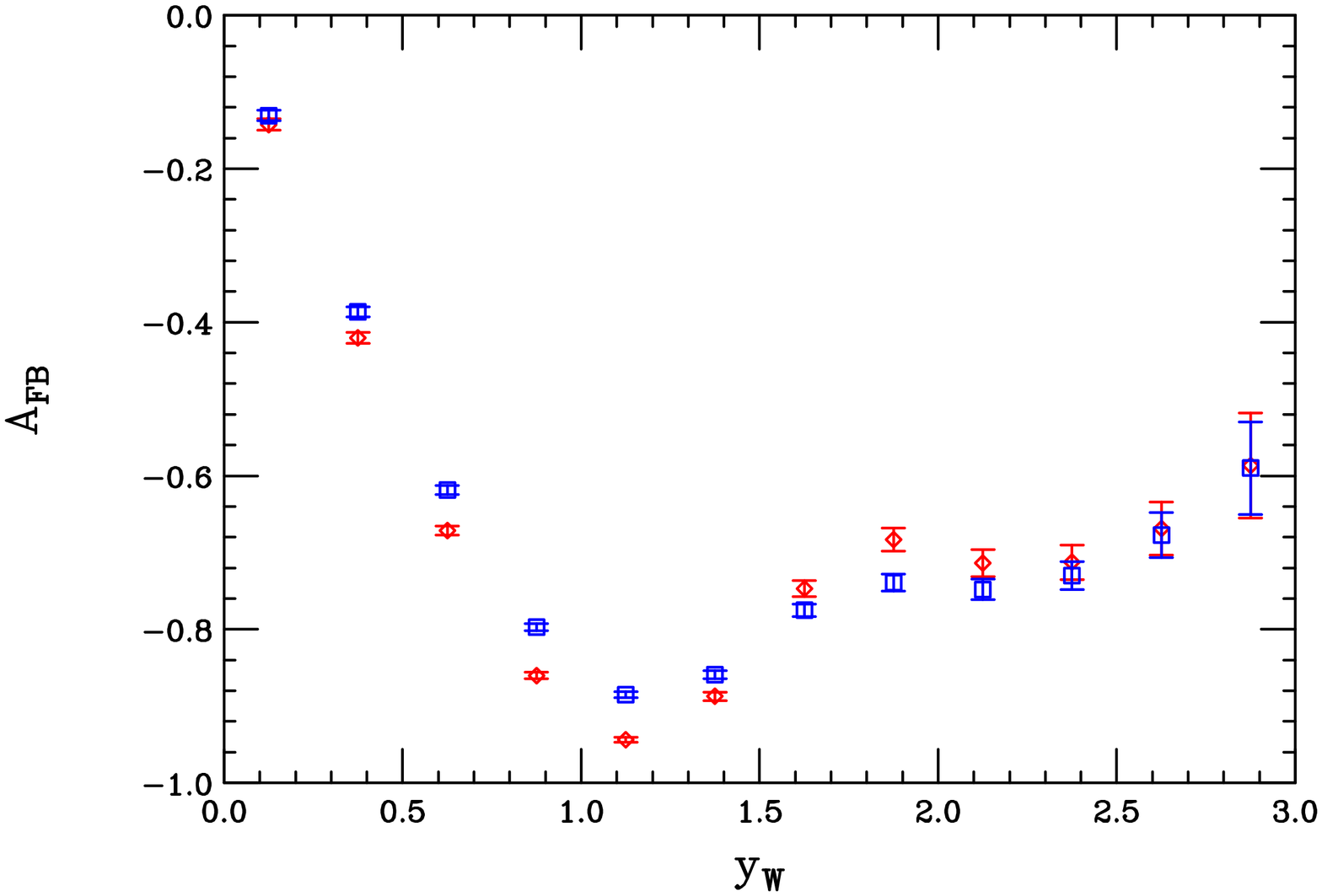}}
\vspace*{0.1cm}
\centerline{
\includegraphics[width=7.5cm,angle=90]{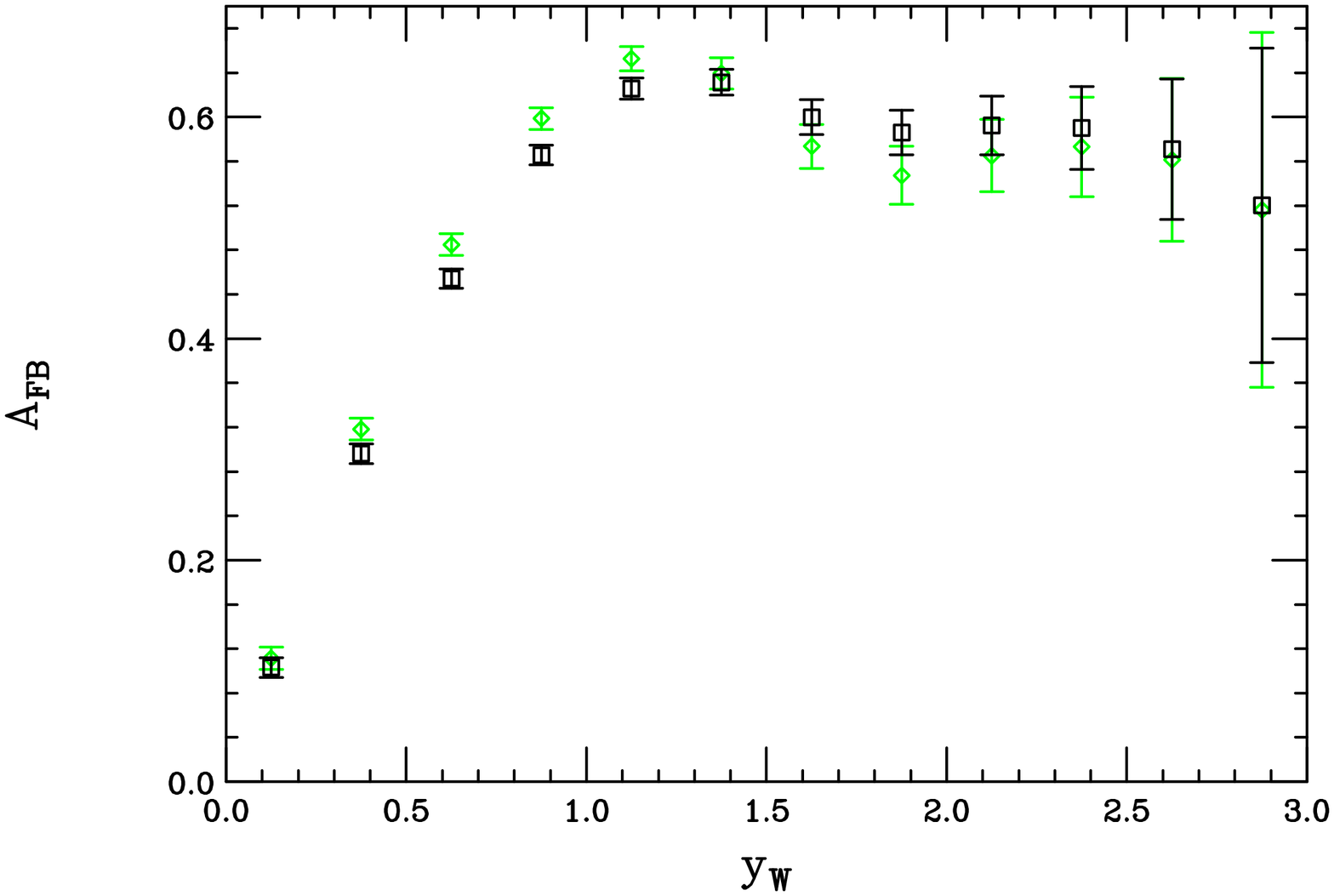}}
\vspace*{0.1cm}
\caption{The value of $A_{FB}$ as a function the center of mass rapidity, $y_W$, integrated over 
the transverse mass bin 400-1000 GeV assuming a mass of 1.5 TeV 
for the $W^{'+}$($W^{'-}$) in the top(bottom) panel. An integrated luminosity of 300 $fb^{-1}$ has been assumed and the distribution has been folded 
around $y_W=0$. The upper(lower) set of data points in the top(lower) panel for small values of $y_W$ corresponds to the choice of $h_{W'}=-1$. Note that 
we have chosen signs to make the ranges of $A_{FB}$ comparable in both cases.}
\label{fig7}
\end{figure}

The next observable we consider is the charge asymmetry, $A_{WQ}(y_W)$:
\begin{equation}
A_{WQ}(y_W)={{N_+(y_W)-N_-(y_W)}\over {N_+(y_W)+N_-(y_W)}}\,,
\end{equation}
where $N_\pm(y_W)$ are the number of events with charged leptons of sign $\pm$ in a given bin of rapidity. Note that at the LHC, $A_{WQ}(y_W)$ is 
symmetric under $y_W \to -y_W$ so that we can again fold the distribution around $y_W=0$. Fig.~\ref{fig8} shows this distribution, integrated over the 
interference region $0.4 \leq M_T \leq 1$ TeV, assuming $M_{W'}=1.5$ TeV and a luminosity of 300 $fb^{-1}$. It is clear 
that at this level of integrated luminosity the two 
distributions are reasonably distinguishable. However, as we lower the luminosity or raise the mass of the $W'$ the quality of the separation degrades 
significantly. Certainly for luminosities less that $\simeq 100$ $fb^{-1}$, this asymmetry measurement would not be very helpful.  Thus $A_{WQ}(y_W)$ 
is not a very useful tool for coupling helicity determination until high luminosities become available. 

\begin{figure}[htbp]
\centerline{
\includegraphics[width=7.5cm,angle=90]{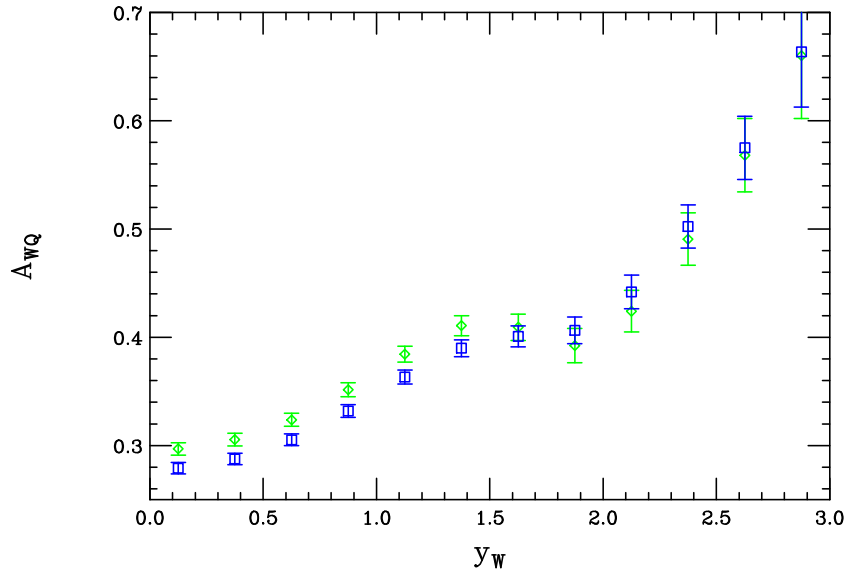}}
\vspace*{0.1cm}
\caption{The $W-W'$ induced charge asymmetry, assuming $M_{W'}=1.5$ TeV, as a function the center of mass rapidity, $y_W$, integrated over 
the transverse mass 
bin 400-1000 GeV. An integrated luminosity of 300 $fb^{-1}$ has been assumed and the distribution has been folded around $y_W=0$. The upper set 
of data points at low values of $y_W$ corresponds to the choice of $h_{W'}=1$.}
\label{fig8}
\end{figure}

A last asymmetry possibility to consider is the rapidity asymmetry for the final state charged leptons themselves,  $A_\ell(y_\ell)$:
\begin{equation}
A_{\ell}(y_\ell)={{N_+(y_\ell)-N_-(y_\ell)}\over {N_+(y_\ell)+N_-(y_\ell)}}\,,
\end{equation}
which is also an even function of $y_\ell$ so the distribution can again be folded around $y_\ell=0$. The resulting distribution can be seen in 
Fig.~\ref{fig9} for large integrated luminosities. Here we again see reasonable model differentiation at low values of $y_\ell \lsim 1$ but this 
fades in utility as integrated luminosities drop much below $\simeq 100$ $fb^{-1}$ as the two curves are generally rather close. 

From this general discussion of possibly asymmetries that one can form employing this final state we can thus conclude that their usefulness in coupling 
helicity determination will require $\simeq 100 fb^{-1}$. 

\begin{figure}[htbp]
\centerline{
\includegraphics[width=7.5cm,angle=90]{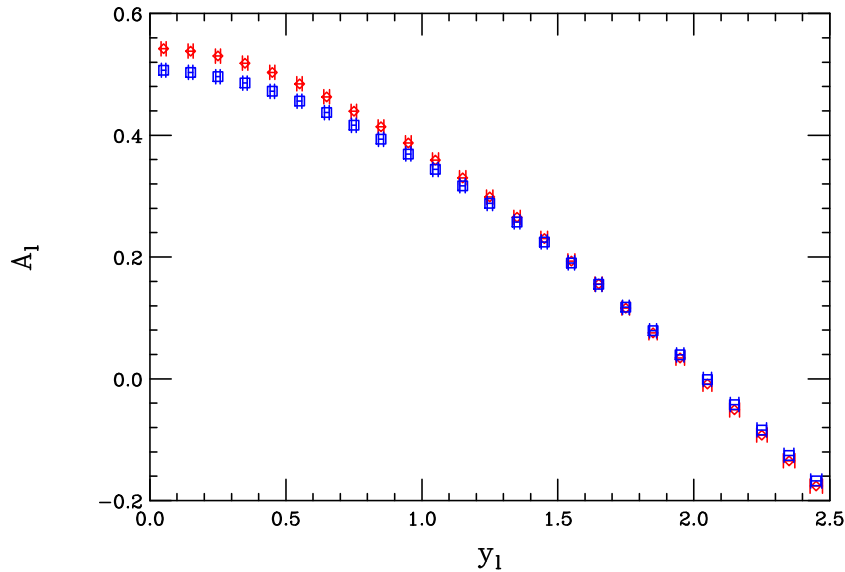}}
\vspace*{0.1cm}
\caption{The $W-W'$ induced lepton asymmetry, assuming $M_{W'}=1.5$ TeV, as a function the lepton's rapidity, $y_\ell$, integrated over 
the transverse mass bin 400-1100 GeV. An integrated luminosity of 300 $fb^{-1}$ has been assumed and the distribution has been folded around 
$y_\ell=0$. The upper set of data points at low values of $y_\ell$ corresponds to the choice of $h_{W'}=1$.}
\label{fig9}
\end{figure}

In the case of extra dimensions we know that an entire tower of $W'$-like KK states is expected to exist. Do the presence of these additional states modify the 
results we have obtained above for an ordinary $W'$? To address this, consider the simplified case of a second $W'$-like KK state which have the same coupling 
strength as the SM $W$ and is twice as heavy as the $W'$ discussed above, \ie,3 TeV. Now imagine that the coupling helicity of this second state is uncorrelated 
with that of the $W'$; in the $M_T$ distribution in the $W-W'$ interference region influenced by this state? The upper panel in Fig. ~\ref{fig99} addresses this 
issue for modest luminosities including the effects of smearing. The upper(lower) set of three histograms corresponds to the case where $h_{W'}=-1(1)$ and either 
there is no $W''$, as above, or $h_{W''}=\pm 1$. This demonstrates that the existence of the extra KK states has little influence on the results we obtained above 
independent of {\it their} coupling helicities.  

Up to now we have assumed that $C_{W'}^\ell=C_{W'}^q$; what if this was no longer true? How would the $M_T$ distribution and our ability to determine coupling 
helicity be modified? The simplest case to examine is $C_{W'}^\ell=-C_{W'}^q=1$ with $h_{W'}=\pm 1$. (Note that interchanging the signs of these two couplings, 
\ie, which one of these two couplings we choose to be negative, has no physical effect on the $M_T$ distribution or on any of the asymmetries discussed earlier.) 
The result of this investigation is shown in the lower panel of Fig. ~\ref{fig99}. Here the red(green) histograms correspond to the cases analyzed above where 
$C_{W'}^\ell=C_{W'}^q=1$ and $h_{W'}=1(-1)$ whereas the blue(magenta) histograms corresponds to the cases where $C_{W'}^\ell=-C_{W'}^q=1$ with $h_{W'}=1(-1)$. 
It is clear from this Figure that the $M_T$ distribution distinguishes only three of these cases with the $C_{W'}^\ell=\pm C_{W'}^q=1, h_{W'}=-1$ possibilities being 
degenerate. The reason for this is that in both these cases there is no interference with the SM $W'$ exchange and in the pure $W'$ term in the cross section this 
sign change is irrelevant; these two degenerate cases are, of course, separable using the information obtained from $A_{FB}$ as they produce values with opposite sign. 

\begin{figure}[htbp]
\centerline{
\includegraphics[width=7.5cm,angle=90]{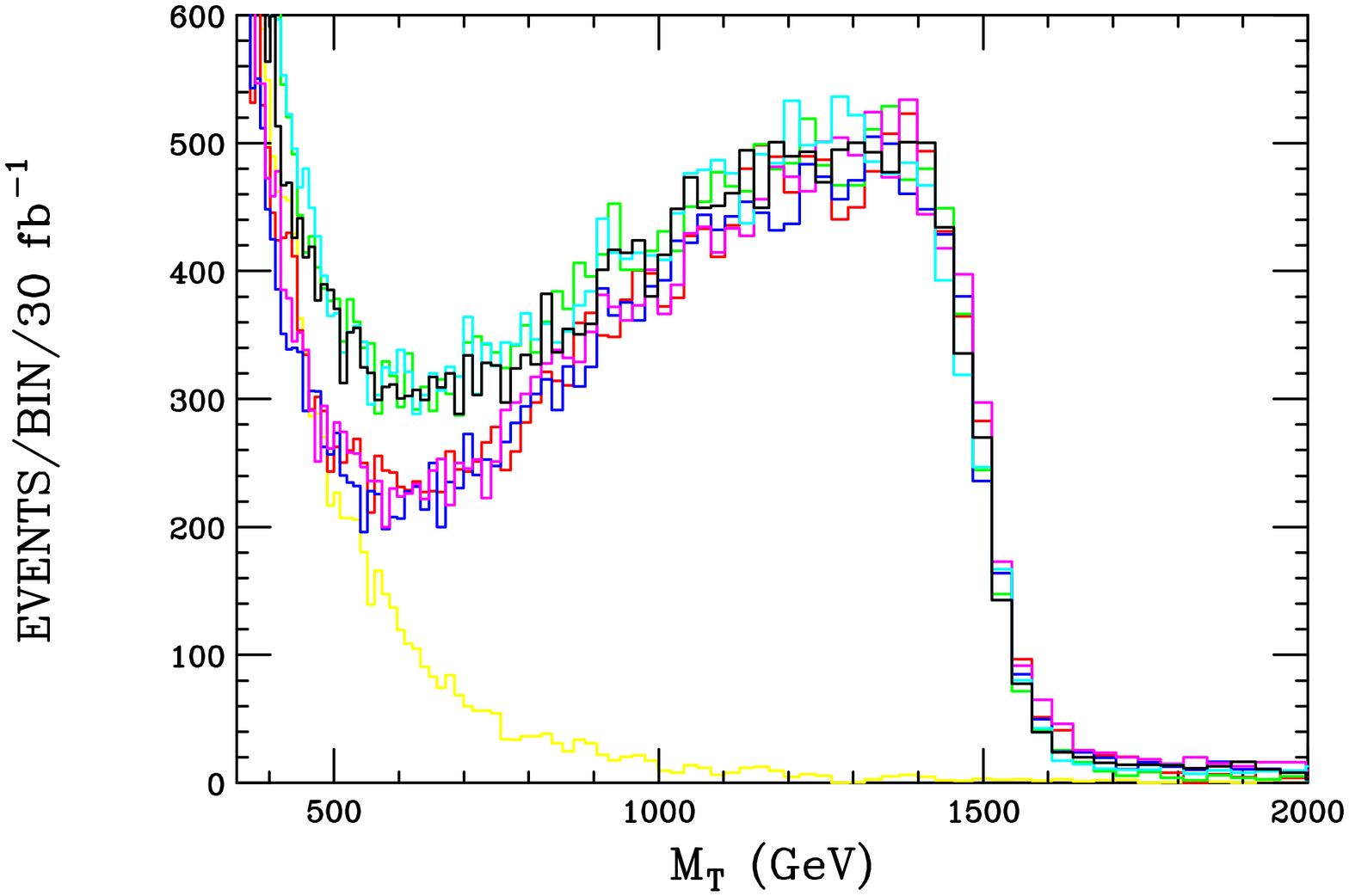}}
\vspace*{0.1cm}
\centerline{
\includegraphics[width=7.5cm,angle=90]{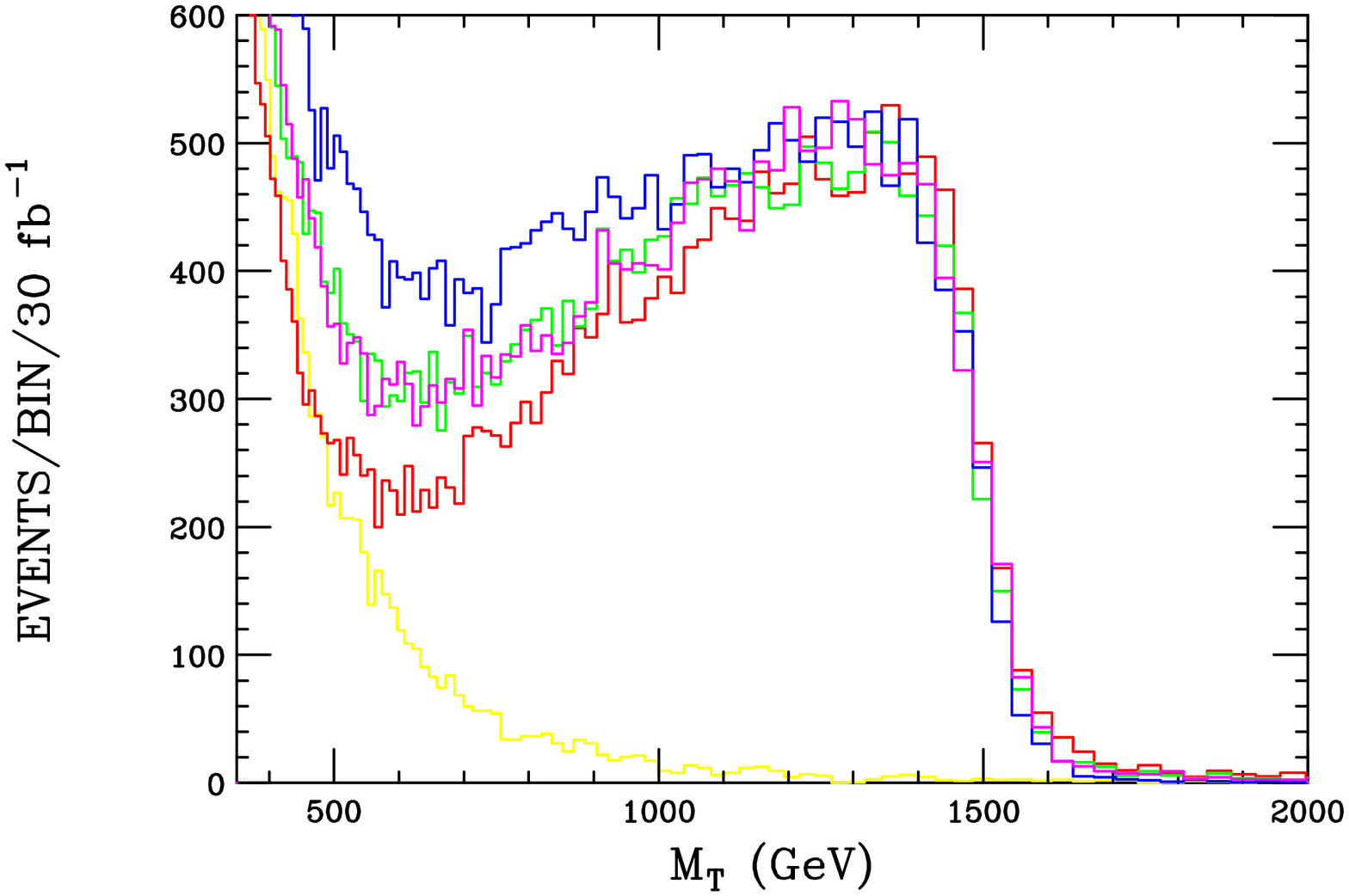}}
\vspace*{0.1cm}
\caption{Smeared $M_T$ distributions for several scenarios; the top panel, the lower(upper) compares the single $W'$ case discussed above to that where a second 
KK state, $W''$, exists with coupling helicities uncorrelated to that of the $W'$. Details are given in the text. In the lower panel, we compare the cases for 
$h_{W'}=\pm 1$ allowing for the possibility that $C_{W'}^\ell=\pm C_{W'}^q$ with the signs uncorrelated with the coupling helicity; the details are 
discussed in the text.}
\label{fig99}
\end{figure}

Lastly, and to be more general, we must at least consider possible scenarios where the couplings of the $W'$ to SM fermions are a substantial admixture of 
both LH and RH helicities, though obvious examples of such kinds of models are apparently absent from the existing literature. To get a feel for such a 
possibility, we perform two analyses: first, we set $C^{\ell,q}=1$ as before and vary the values of  $h_{W'}$ between pairs of positive and negative values. 
As we do this, the helicity of the 
couplings of the $W'$ will vary as will its total decay width which behaves as $\sim 1+h_W^2$. In a second analysis, we can rescale the 
values of the $C^{\ell,q}$ so that the $W'$ width is held fixed.  In this case, as we will see, the resulting histograms for the transverse mass 
distribution lie especially close to one another. The results of these two sets of calculations are shown in Fig.~\ref{fig10} in the case of large integrated 
luminosities assuming the default value of $M_{W'}=1.5$ TeV. 
In the first analysis shown in the top panel, we see that at these assumed luminosities all of the different histograms 
are distinguishable and not just the two pairs of cases with opposite helicities. This result generally remains true down to luminosities $\sim 75 fb^{-1}$ 
or so. If we are {\it only} interested in separating opposite helicity pairs then we find that the cases $h_{W'}=\pm 0.8(0.6,0.4,0.2)$ can be distinguished 
down to luminosities of order $\sim 10(25,50,75) fb^{-1}$, respectively. 

In the second analysis, as seen in the lower panel of the figure, the histograms for $h_{W'}=0.8,~0.6$ and 0.4 (as well as for 
their corresponding opposite helicity partners) are very close to one another and are essentially inseparable even at these high 
luminosities. However, the two sets of opposite helicity histograms remain distinguishable and this will remains true down to luminosities 
of order $30-75~fb^{-1}$. It would seem from these analyses that the transverse mass distribution will play the dominant role in $W'$ 
coupling helicity determination in all possible cases although somewhat higher integrated luminosities may be required in some scenarios.

\begin{figure}[htbp]
\centerline{
\includegraphics[width=7.5cm,angle=90]{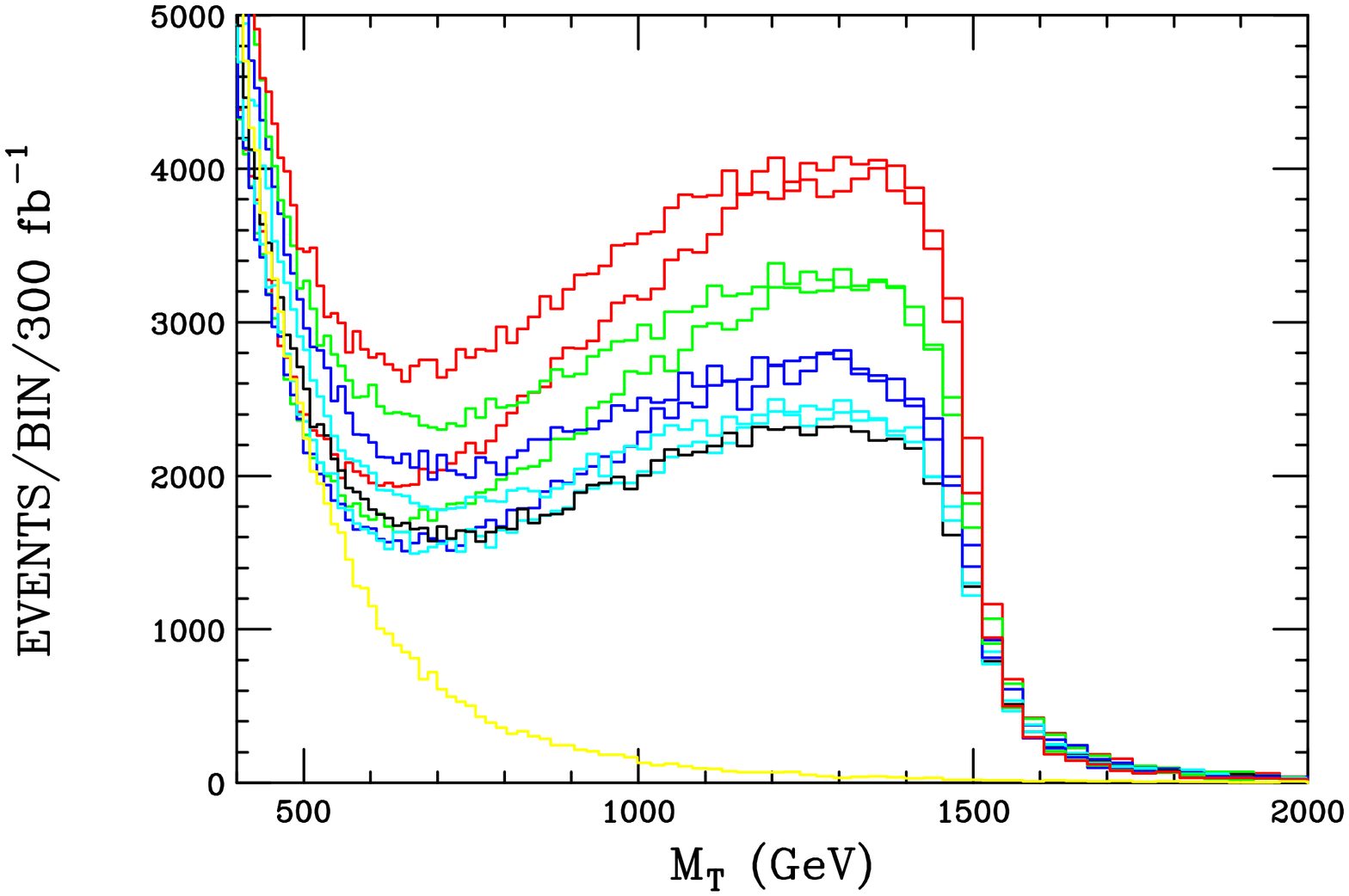}}
\vspace*{0.1cm}
\centerline{
\includegraphics[width=7.5cm,angle=90]{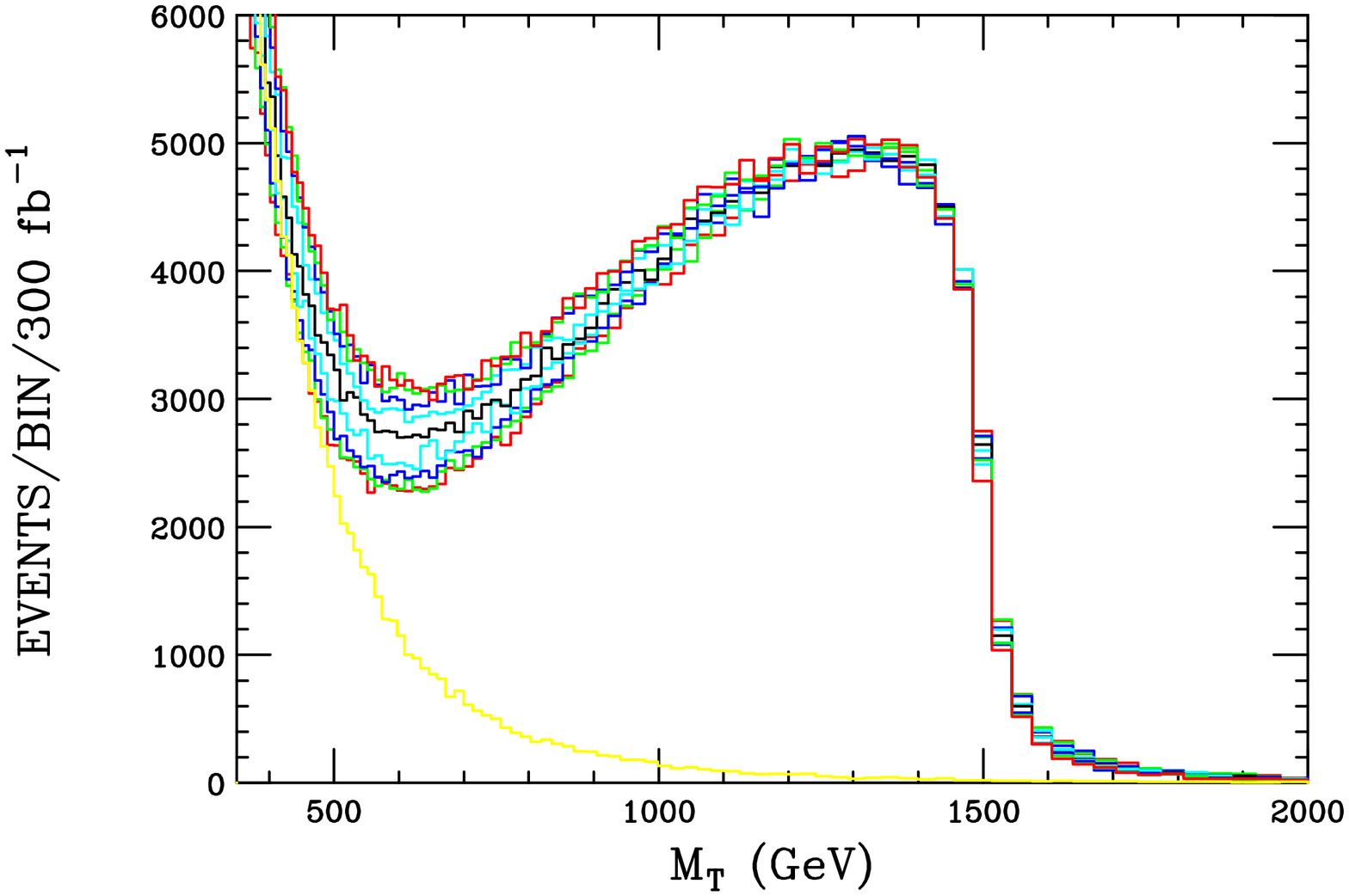}}
\vspace*{0.1cm}
\caption{Same as the linear plot shown in Fig.1, but now for other values of the coupling helicities. From top to bottom the 
pairs of histograms in the upper panel correspond to $h(W')=\pm 0.8,\pm 0.6,\pm 0.4 and \pm 0.2$, respectively. The next lowest 
single histogram corresponds to the case of pure vector couplings, \ie, $h(W')=0$. In producing these results we have assumed that 
the values of the $C^{\ell,q}$=1. In the lower panel, we show the same result now but with the overall couplings rescaled so as to 
keep the $W'$ width a constant.}
\label{fig10}
\end{figure}

\section{Summary and Conclusions}

Apart from its mass and width, the most important property of a new charged gauge boson, $W'$, is the helicity of its couplings to the SM fermions. Such 
particles are predicted to exist in the TeV mass range in many new physics models and this coupling helicity is an order one discriminator between the 
various classes of models. The main difficulties with the existing techniques for determining this helicity are potentially threefold: 
($i$) they require rather high integrated luminosities even for a relatively light $W'$, and/or ($ii$) 
they are sufficiently intricate as to require a detailed background and detector study to determine their feasibility, and/or ($iii$) they make use of 
more complex final states other than the standard $\ell +E_T^{miss}$ discovery channel. Some of these techniques also suffer from employing the narrow 
width approximation which can result in loss of valuable information regarding the effects of $W-W'$ interference. 
In this paper we propose a simple technique for making this helicity determination at the LHC. In order to attempt to circumvent all 
of these difficulties, we have examined the $W-W'$ interference region of the transverse mass distribution for the $\ell +E_T^{miss}$ discovery mode. We 
have found that this distribution is particularly sensitive to the helicity of the $W'$ couplings. In particular, using this technique we have shown that 
such helicity differentiation requires only $\sim 10(60,300)~fb^{-1}$ assuming $M_{W'}=1.5(2.5,3.5)$ TeV and provided that the $W'$ has Standard Model strength 
couplings. This helicity determination can be further strengthened by the use of various discovery channel leptonic asymmetries also measured in the same 
interference regime once higher integrated luminosities are available as well as by the more traditional approaches. Hopefully the LHC will observe a $W'$ 
so that this approach can be employed.

\noindent{\Large\bf Acknowledgments}

The author would like to thank A. De Roeck, S.Godfrey and J. Hewett for input and discussions related to this paper.

%
\def\MPL #1 #2 #3 {Mod. Phys. Lett. {\bf#1},\ #2 (#3)}
\def\NPB #1 #2 #3 {Nucl. Phys. {\bf#1},\ #2 (#3)}
\def\PLB #1 #2 #3 {Phys. Lett. {\bf#1},\ #2 (#3)}
\def\PR #1 #2 #3 {Phys. Rep. {\bf#1},\ #2 (#3)}
\def\PRD #1 #2 #3 {Phys. Rev. {\bf#1},\ #2 (#3)}
\def\PRL #1 #2 #3 {Phys. Rev. Lett. {\bf#1},\ #2 (#3)}
\def\RMP #1 #2 #3 {Rev. Mod. Phys. {\bf#1},\ #2 (#3)}
\def\NIM #1 #2 #3 {Nuc. Inst. Meth. {\bf#1},\ #2 (#3)}
\def\ZPC #1 #2 #3 {Z. Phys. {\bf#1},\ #2 (#3)}
\def\EJPC #1 #2 #3 {E. Phys. J. {\bf#1},\ #2 (#3)}
\def\IJMP #1 #2 #3 {Int. J. Mod. Phys. {\bf#1},\ #2 (#3)}
\def\JHEP #1 #2 #3 {J. High En. Phys. {\bf#1},\ #2 (#3)}

\end{document}